\documentclass[prx,aps,superscriptaddress,twocolumn,notitlepage,floatfix,10pt]{revtex4-2}
\usepackage{amsmath,amssymb}
\usepackage[american]{babel}
\usepackage{graphicx}
\usepackage[colorlinks,
citecolor=red,
linkcolor=red,
urlcolor=red]{hyperref}
\usepackage{orcidlink}
\newcommand{\Tr}{\operatorname{Tr}}

\newcommand{\ket}[1]{|{#1}\rangle}
\newcommand{\bra}[1]{\langle {#1}|}

\newcommand{\titleinfo}{Theory of Magic Phase Transitions in Encoding-Decoding Circuits}

\begin{document}
\title{\titleinfo}

\author{Piotr Sierant~\orcidlink{0000-0001-9219-7274}}
\email{piotr.sierant@bsc.es}
\affiliation{Barcelona Supercomputing Center Plaça Eusebi G\"uell, 1-3 08034, Barcelona, Spain}

\author{Xhek Turkeshi~\orcidlink{0000-0003-1093-3771}}
\email{turkeshi@thp.uni-koeln.de}
\affiliation{Institut f\"ur Theoretische Physik, Universit\"at zu K\"oln, Z\"ulpicher Strasse 77, 50937 K\"oln, Germany}

\date{\today}

\begin{abstract}
Quantum magic resources, or nonstabilizerness, are a central ingredient for universal quantum computation. In noisy many-body systems, the interplay between these resources and errors leads to sharp magic phase transitions. However, the microscopic mechanism behind these critical phenomena is still an open question, especially since early empirical evidence showed conflicting results regarding their universality classes. 
In this work, we provide a comprehensive picture of magic phase transitions for the class of encoding-decoding quantum circuits to resolve these ambiguities. 
We analytically show that the behavior of magic resources is fundamentally dictated by the chosen measurement protocol. When we fix, or post-select, the class of measurement syndromes, the magic transition inherits the universal features of the error-resilience phase transition in the circuits. Interestingly, this clean transition survives even for fully random Haar encoders showing that it is a consequence of initial's state retrieval, and not an artifact of the Clifford encoders.
On the other hand, if we consider realistic Born-rule sampling, the intrinsic statistical fluctuations of a given syndrome measurement act as a relevant perturbation. 
This brings in strong finite-size drifts and an apparent multifractality, which end up altering the critical behavior of the system.
We reveal that magic phase transitions are actually direct manifestations of error-resilience thresholds, rather than independent critical phenomena, reconciling conflicting observations from the earlier literature. Ultimately, our framework clarifies how the quantum computational power can survive, or be irreversibly destroyed, due to the competition between scrambling, measurements, and errors.
\end{abstract}
\maketitle

\section{Introduction}

The interplay between intrinsic unitary evolution in many-body systems and the dissipative action of environmental probes~\cite{fisher2023randomquantumcircuits, bertini2026nonequ} has revealed unique dynamical phases of matter with no analogue in equilibrium statistical mechanics. 
Unlike traditional condensed matter systems, these synthetic phases are not characterized by local order parameters, such as magnetization, but rather by the system's capacity to process, propagate, and preserve quantum information. 
\textit{Mutatis mutandis}, these phases can be rigorously understood through their robustness against quantum errors. 
In this viewpoint, the system's resilience to noise is encoded in nonlocal structural properties, such as entanglement or higher-order correlations, rather than local observables.

A paradigmatic example is the measurement-induced phase transition (MIPT) in monitored quantum circuits~\cite{li2018quantumzenoeffect,skinner2019measurementinducedphase,fisher2023randomquantumcircuits, chan2019unitaryprojective, potter2022quantumsciencesandtechnology}.
In these systems, the dynamics are governed by a competition between two opposing forces: unitary scrambling, which acts as an encoding map that hides information in nonlocal correlations, and projective measurements, which act as errors that extract information and disentangle the state.
Varying the measurement rate drives a sharp transition~\cite{li2019measurementdrivenentanglement,zabalo202criticalpropertiesof,sierant2022dissipativefloquetdynamics}.
Below a critical threshold, the system exhibits an error-resilient phase in which logical information is protected; beyond this threshold, it enters an error-vulnerable phase in which the encoding capacity collapses~\cite{gullans2020dynamicalpurificationphase,gullans2020scalableprobesof,choi2020quantumerrorcorrection}.
Often, the critical point displays universal features governed by conformal field theory and distinct from standard unitarity classes~\cite{jian2020measurementinducedcriticality,bao2020theoryofthe,sierant2022measurementinducedphase,sierant2023controllingentangleat, Nahum23reno}.

This error-correcting perspective provides deep insights into increasingly complex problems, ranging from the inclusion of incoherent noise~\cite{turkeshi2024error,lovaz2024quantum} to complexity phase transitions in random circuit sampling~\cite{morvan2023phase,ghosh2023complexity,ware2023asharpphase}.
Recently, this framework has been extended to \textit{magic resources}, i.e., non-stabilizerness, a crucial resource for universal quantum computation~\cite{bravyi2005universal}.
Magic resources are the ``missing ingredient'' that elevate Clifford circuits, which are fault-tolerantly implementable but classically simulatable, into universal quantum computation paradigms~\cite{ chitambar2019quantum,liu2022manybody}.
In monitored dynamics, a sharp transition in magic density can emerge from a competition between two mechanisms: coherent non-Clifford gates that inject magic resources~\cite{Haferkamp2022,zhou2020single,Leone2024LearningTDoped,Chia2024SingleCopyTDoped,Gu2024DopedStabilizerManyBody,mao2025qudit,szombathy2025spectralpropertiesversusmagic,magni2025anticoncentration,leone2025noncliffordcostrandomunitaries,Zhang2025DesignsMagicAugClifford,liu2025classicalsimulabilitycliffordtcircuits,nakhl2025stabilizer,Vairogs2025,scocco2025risefallnonstabilizernessrandom,lóio2026quantumstatedesignsmagic,huang2025advantageutilizingnonlocalmagic}, driving the system away from the stabilizer manifold~\cite{Leone2021quantumchaosis,Magni2025quantumcomplexity,Haug2025probingquantum,Iannotti_2026,Turkeshi_2024_2,varikuti2025impactcliffordoperationsnonstabilizing,santra2025complexitytransitionschaoticquantum,hou2025stabilizerentanglementenhancesmagic,dowling2025magic,Faidon,aditya2025mpembaeffectsquantumcomplexity,aditya2025growthspreadingquantumresources,zhang2024quantummagicdynamicsrandom,varikuti2025deepthermalizationmeasurementsquantum,maity2025localspreadingstabilizerrenyi,bejan2025magicspreadingunitaryclifford}, and Pauli measurements that suppress nonstabilizerness by projecting the state back toward the stabilizer manifold~\cite{Niroula2024phasetransition}.

Despite these intuitive arguments, early empirical evidence regarding magic phase transitions raised fundamental questions.
The current literature reports conflicting universality classes, drifting critical points, and discrepancies between different magic monotones.
For example, the seminal work by Niroula \textit{et al.}~\cite{Niroula2024phasetransition} reported magic transitions in random Clifford circuits interspersed with non-Clifford gates, but extracted critical exponents that were difficult to reconcile within a single universality class.
It therefore remains debated whether magic transitions represent a distinct critical phenomenon or merely a shadow of a deeper underlying entanglement transition~\cite{bejan2024dynamical,fux2024entanglement,paviglianiti2024estimatingnonstabilizernessdynamicssimulating,aziz2025classicalsimulationslowmagic,tirrito2025magicphasetransitionsmonitored,wang2025magictransitionmonitoredfree,Tarabunga2025magic}.

\begin{figure}[t!]
    \centering
    \includegraphics[width=1\linewidth]{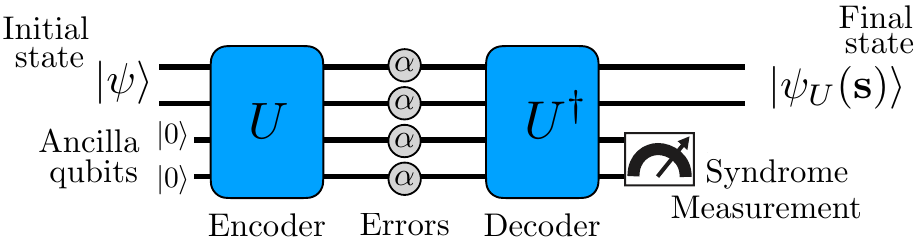}
 \caption{\textbf{Encoding--decoding circuit}. The unitary $U$ encodes a $k$-qubit logical state into the $N$-qubit code space. The circuit comprises the encoder $U$, a layer of local errors $E_i$, and the decoder $U^\dagger$. The decoded logical state $|\psi_U(\mathbf{s})\rangle$ is obtained by projecting the ancilla qubits onto $\ket{\mathbf{s}}$.}
    \label{fig:cartoon}
\end{figure}

In this work, we present a comprehensive theory of magic phase transitions in the family of encoding--decoding circuits~\cite{turkeshi2024error,dallas2025butterflyeffectencodingdecodingquantum,dallas2025nonlocalmagicgenerationinformation,sommers2025spectralpropertiescodingtransitions,Khindanov2025,Chen2025subsystem,behrends2025the,cheng2025emergent}.
Operationally, this framework proceeds in three stages [cf. Fig.~\ref{fig:cartoon}]: (i) a logical state is mapped into a nonlocal code subspace via a unitary encoder; (ii) the system is subjected to a layer of coherent (magic-injecting) or incoherent errors; and (iii) the state is decoded, followed by syndrome measurements that attempt to restore the logical information.

To pin down the mechanism behind the magic transition, we combine exact ensemble averages based on Weingarten calculus with large-scale numerical simulations.
This framework resolves ambiguities in the literature by cleanly separating two measurement protocols:
\begin{itemize}
    \item \textbf{Forced Measurements (Fixed Syndrome Class):} We show that when the decoding syndrome is fixed (postselected) to a specific outcome (e.g., the trivial syndrome $\mathbf{s}=\mathbf{0}$), the transition is driven purely by the competition between the error strength and the code distance.
    In this regime, we derive exact analytical bounds for the Stabilizer Rényi Entropy (SRE)~\cite{Leone_2022,Leone_2024,cusumano2025nonstabilizernessviolationschshinequalities,bittel2025operationalinterpretationstabilizerentropy} and the fidelity, showing that they act as order parameters for the error-correcting phase.
    Crucially, we demonstrate that a magic transition exists even for Haar-random circuits in this regime, a setting typically associated with volume-law entanglement and magic resources~\cite{turkeshi2025pauli,Iannotti2025entanglement,szombathy2025independentstabilizerrenyientropy}, proving that the transition is a fundamental feature of error correction, rather than an artifact of Clifford distinctiveness.

    \item \textbf{Born-rule Measurements (Sampled Syndrome):} We show that the ``magic transitions'' observed in previous experiments correspond to averaging over syndromes weighted by their Born probabilities $p(\mathbf{s})$~\cite{Niroula2024phasetransition}.
    We demonstrate that the probability distribution of syndromes acts as a relevant perturbation to the error-resilience fixed point.
    At finite system sizes, a competition arises between the intrinsic error-correction threshold and statistical fluctuations of syndrome outcomes.
\end{itemize}

Disentangling these two protocols clarifies how magic dynamics are controlled by error correction, thus resolving open challenges in the literature.
In the forced ensemble, the microscopic origin of the magic transition is an \textit{error-resilience phase transition}~\cite{turkeshi2024error}, whereas Born-rule sampling introduces a relevant perturbation through the syndrome statistics.
This distinction provides a unified interpretation of the observed multifractal phenomenology and finite-size drifts in encoding--decoding circuits.

The remainder of this paper is organized as follows.
In \textbf{Sec.}~\ref{sec:methods}, we establish the theoretical framework for encoding--decoding circuits. We define the models for Haar and Clifford encoders, introduce the key quantities of interest (fidelity and stabilizer R\'enyi entropy), and detail the averaging techniques used to derive our results.
\textbf{Sec.}~\ref{sec:forced} focuses on ``forced'' measurements. We analyze the error-resilience transition in the fixed-syndrome ensemble, present exact analytical results for the Haar case, and demonstrate the universality of the transition in the Clifford case.
In \textbf{Sec.}~\ref{sec:born}, we turn to ``Born-rule'' measurements. We first address the implications for Haar circuits and then focus on the Clifford case, showing how syndrome sampling acts as a relevant perturbation that modifies the critical behavior.
Finally, we discuss the broader implications of these findings in \textbf{Sec.}~\ref{sec:discussion} and conclude in \textbf{Sec.}~\ref{sec:conclusion}.
Additional analysis and details are relegated to the \textbf{Appendices}.

\section{Encoding-decoding circuits}
\label{sec:methods}

In this section, we establish the theoretical framework used to investigate magic phase transitions.
We focus on the architecture of encoding--decoding circuits, which serves as a minimal model for the competition between information scrambling and error accumulation.
We analyze Haar-random circuits to derive exact analytical bounds in the thermodynamic limit, and we study random Clifford circuits to probe the interplay between magic resources and stabilizer structure.

We begin by defining the circuit protocol and the error model under consideration. We then introduce the figures of merit---specifically, the fidelity and the stabilizer R\'enyi entropy~\cite{Leone_2022}---which serve as order parameters for the magic transition. Finally, we detail the averaging procedures employed to qunatify the transitions in encoding-decoding circuits.

\subsection{The circuit architecture}
\label{subsec:model}
We study a system of $N$ physical qubits with Hilbert space $\mathcal{H}$, partitioned into a logical subsystem $L$ of $k$ qubits and a syndrome subsystem $S$ of $N-k$ ancilla qubits, defining a code rate $r=k/N$.
The dynamics implement a three-step encoding--decoding protocol~\cite{turkeshi2024error}.
First, the system is initialized in the pure state $\rho_0=|\Psi_0\rangle\langle \Psi_0|$, with $|\Psi_0\rangle\equiv |\psi\rangle_L \otimes |\mathbf{0}\rangle_S$, where $|\psi\rangle_L$ is an arbitrary logical state and $|\mathbf{0}\rangle_S=|0\rangle^{\otimes (N-k)}$ denotes the reference ancilla state.
Throughout this work, we often specialize to $|\psi\rangle_L=|\mathbf{0}\rangle_L\equiv |0\rangle^{\otimes k}$. 
An $N$-qubit unitary encoder $U$ embeds this product state into the $2^N$-dimensional code space, scrambling the logical information across all qubits.

Subsequently, the encoded state is subjected to local errors
$E = E_N\circ E_{N-1}\circ \dots \circ E_1$, where $\circ$ denotes composition and $E_j$ acts on site $j$.
While our formalism accommodates general channels, we focus on coherent errors modeled as unitary channels generated by a layer of single-qubit $Z$ rotations $E_j(\rho)=e^{-i \frac{\alpha}{2} Z_j} \rho e^{i \frac{\alpha}{2} Z_j}$, with $\alpha\in[0,\pi]$ setting the error strength~\cite{Bravyi_2018,behrends2022surface,zhu2023nishimori,lee2025mixed}. For later convenience, we define 
\begin{equation}
    V_{\alpha} = \bigotimes_{j=1}^N e^{-i \frac{\alpha}{2} Z_j},
    \label{eq:coherent_error}
\end{equation}
so that the full error layer is given by $E(\rho) = V_{\alpha}\rho V_{\alpha}^\dagger$.

In the final stage, we decode with $U^\dagger$, obtaining $\rho_U = U^\dagger E\bigl(U \rho_0 U^\dagger\bigr) U$, and perform a projective measurement on the syndrome subsystem $S$.
For a syndrome bitstring $\mathbf{s}\in\mathbb{Z}_2^{N-k}$, we introduce the projector $\Pi_{I,\mathbf{s}} = I_L\otimes \ket{\mathbf{s}}\!\bra{\mathbf{s}}_S$, where $I_L$ is the identity operator on the logical subspace. 
With this operator, the partial trace over the syndrome space reads $\Tr_S(A)=\sum_{\mathbf{s}} \Tr(\Pi_{I,\mathbf{s}}A)$. 
Conditioned on the outcome $\mathbf{s}$, the logical post-measurement state is
\begin{equation}
\rho_{L,U}(\mathbf{s})=\frac{1}{\Tr[\Pi_{I,\mathbf{s}}\rho_U]}\,
\Tr_S\!\left[\Pi_{I,\mathbf{s}}\rho_U\Pi_{I,\mathbf{s}}\right],
\label{eq:rhoL}
\end{equation}
which occurs with Born probability $p_{U}(\mathbf{s})=\Tr[\Pi_{I,\mathbf{s}}\rho_U]$.

For coherent noise $V_\alpha$ in Eq.~\eqref{eq:coherent_error} and the pure input state $\rho_0$, the conditioned logical state is pure and can be written as
\begin{equation}
\ket{\psi_{U}(\mathbf{s})}_L=
\frac{\bra{\mathbf{s}}_S\,U^\dagger V_{\alpha}U\,\ket{\Psi_0}}
{\bigl\|\bra{\mathbf{s}}_S\,U^\dagger V_{\alpha}U\,\ket{\Psi_0}\bigr\|}.
\label{eq:stateS1}
\end{equation}
In the following, we occasionally drop the $L$ and $S$ subscripts when they are clear from context.

To expose the interplay between error correction and magic, we consider two complementary ensembles $\mathcal{E}$ for the encoder $U$: (i) Haar-random encoders, with $U$ drawn from the Haar measure on the unitary group $\mathcal{U}(2^N)$, admit an exact analytical treatment via Weingarten calculus~\cite{turkeshi2024error}; (ii) random Clifford encoders, sampled uniformly from the $N$-qubit Clifford group $\mathcal{C}_N$, mirror the experimental setting of Ref.~\cite{Niroula2024phasetransition}. In the Clifford case, the encoder itself generates no magic; consequently, any non-stabilizerness in the logical output $\rho_{L,U}(\mathbf{s})$ is induced solely by the coherent layer $V_\alpha$.

\subsection{Quantities of interest}
\label{subsec:quantities}
To map out the phase diagram and diagnose the magic transition in encoding--decoding circuits, we track two complementary observables: (i) the recovery fidelity of the decoded logical state and (ii) the stabilizer R\'enyi entropy (SRE), which quantifies non-stabilizerness (magic resources) for pure qubit states.

\paragraph{Fidelity.}
The fidelity serves as an order parameter for the error-protecting and error-vulnerable phases in encoding--decoding protocols~\cite{turkeshi2024error}.
Given the conditional logical state $\rho_{L,U}(\mathbf{s})$ in Eq.~\eqref{eq:rhoL}, we quantify recovery by the overlap with the initial logical input $\ket{\psi}$, i.e.,
\begin{equation}
F_U(\mathbf{s})
\equiv
\bra{\psi}\rho_{L,U}(\mathbf{s})\ket{\psi}.
\label{eq:Fs}
\end{equation}
In encoding--decoding circuits, the average fidelity $\overline{F}$ sharply distinguishes an error-protecting phase ($\overline{F}\to 1$) from an error-vulnerable phase ($\overline{F}\to 0$) in the limit $k,N\to\infty$ at fixed code rate $r$.

\paragraph{Stabilizer R\'enyi entropy.}
The central figure of merit in this work is the amount of non-stabilizerness (magic resources) that survives in the logical state. We quantify it using the stabilizer R\'enyi entropy (SRE) of index $q$, defined~\cite{Leone_2022} as
\begin{equation}
    M_q(\rho_{L,U}) = \frac{1}{1-q} \log_2 \sum_{P \in \mathcal{P}_k} \frac{\mathrm{Tr}(\rho_{L,U} P)^{2q}}{2^k},
    \label{eq:sreDEF}
\end{equation}
where $\mathcal{P}_k$ denotes the group of Pauli strings on $k$ qubits. The SRE is a faithful measure of magic: $M_q(\rho)=0$ if and only if $\rho$ is a stabilizer state, and it is strictly positive otherwise. Moreover, it is invariant under Clifford unitaries, additive under tensor products, and non-increasing under pure-state stabilizer protocols~\cite{Leone_2024}; see also Ref.~\cite{Haug23monotones}.

The behavior of $M_q$ reveals whether the magic injected by the coherent errors (and, in the Haar case, by the encoding--decoding unitaries) is purged by the syndrome measurement or instead propagates into the logical subspace.
We focus in particular on the second-order stabilizer R\'enyi entropy $M_2$,
for which we derive rigorous analytical expressions using an eight-replica
calculation for both Haar and Clifford encoders.

\subsection{Our approach}

\subsubsection{Quenched and annealed averages}
To study magic transitions in encoding--decoding circuits, we combine analytical and numerical techniques to compute \textit{quenched} and \textit{annealed} averages of the quantities of interest over realizations of the encoding unitary $U$. We denote by $\Lambda_U(\mathbf{s})$ the value of an observable ($\Lambda=F$ for the fidelity or $\Lambda=M_q$ for the SRE) for a fixed encoder $U$ and syndrome outcome $\mathbf{s}$.
In general, these observables can be written as a function $\Lambda(\mathbf{s})= f(m(\mathbf{s})/p(\mathbf{s}))$ for some polynomials $m(\mathbf{s})$ and $p(\mathbf{s})$ in the unitary encoder $U$.
The quenched average at fixed $\mathbf{s}$ is
\begin{equation}
\overline{\Lambda}(\mathbf{s}) = \mathbb{E}_{U\sim \mathcal{E}}\!\left[ \Lambda_U(\mathbf{s}) \right] = \mathbb{E}_{U\sim \mathcal{E}}\!\left[ f\left(\frac{m(\mathbf{s})}{p(\mathbf{s})}\right)\right],
\label{eq:lam_quench}
\end{equation}
where $\mathbb{E}_{U\sim \mathcal{E}}$ denotes the average over the ensemble $\mathcal{E}$, i.e., over the unitary group $\mathcal{U}(2^N)$ or the Clifford group $\mathcal{C}_N$.
The annealed average $\tilde{\Lambda}_U(\mathbf{s})$, in contrast, is given by
\begin{equation}
\tilde{\Lambda}(\mathbf{s}) \equiv f\left(\frac{\mathbb{E}_{U\sim \mathcal{E}}[m(\mathbf{s})]}{\mathbb{E}_{U\sim \mathcal{E}}[p(\mathbf{s})]}\right)\;.
\label{eq:lam_ann}
\end{equation}
The averages over the polynomials $m$ and $p$ can be evaluated analytically in the replica space $\mathcal{H}^{\otimes n}$, where $n>1$ is an integer.
We compute exact expressions for the annealed averages $\tilde{\Lambda}_U(\mathbf{s})$ using Weingarten calculus~\cite{Collins2006,Roberts17,Mele2024introductiontohaar} and the structure of the Clifford commutant~\cite{Gross2021,magni2025anticoncentration,Magni2025quantumcomplexity,Bittel25commutant,magni2025anticoncentrationstatedesigndoped}.
The exact numerical evaluation of the stabilizer entropy is implemented with the state-of-the-art algorithms~\cite{sierant2026computingquantummagicstate,huang2026fastexactapproachstabilizer,xiao2026exponentiallyacceleratedsamplingpauli}

In general, the annealed and quenched averages differ, $\tilde{\Lambda}(\mathbf{s}) \neq \overline{\Lambda}(\mathbf{s})$.
However, as we will show, in encoding--decoding circuits with forced measurements $\mathbf{s}=\mathbf{0}$, the two averages coincide in the thermodynamic limit due to asymptotic self-averaging,
\begin{equation}
\tilde{\Lambda}(\mathbf{0}) \stackrel{N \to \infty}{\longrightarrow} \overline{\Lambda}(\mathbf{0}).
\label{eq:selfave}
\end{equation}
Beyond this analytically treatable framework, we will study numerically the problem when $\mathbf{s}\neq\mathbf{0}$ and when Born sampling is performed. 
Specifically, for Haar-random encoders we perform full state-vector simulations to estimate quenched averages of the quantities of interest for systems of up to $N=24$ qubits. For Clifford encoders, we instead use the Pauli-propagation method described in the next subsection.

\subsubsection{Pauli-propagation method for Clifford encoders}
\label{sec:PauliPropaCliff}
For Clifford encoders, the structure of the decoded logical state $\ket{\psi_{U}(\mathbf{s})}_L$ is particularly simple. This allows us to develop a numerical technique in the spirit of Pauli-propagation methods~\cite{Rall19,Rudolph25,Loizeau25,Begusic25,GonzalezGarcia25}. While such methods typically track time-evolved operators as superpositions of Pauli strings in the Heisenberg picture, we adapt this framework to the Schr\"{o}dinger picture to directly compute the time-evolved many-body state. Measuring the ancilla qubits projects the system from a $2^N$-dimensional code space into a much smaller $2^k$-dimensional logical subspace. Consequently, our approach yields the numerically exact state vector of the decoded logical state without ever constructing the full code-space state, enabling us to simulate system sizes $N$ well beyond the reach of brute-force methods.

The operator $U^\dagger V_{\alpha} U$ entering Eq.~\eqref{eq:stateS1} simplifies for $U=C\in\mathcal{C}_N$ because Clifford unitaries map Pauli operators to Pauli operators (up to a phase $\pm 1$). Defining
\begin{equation}
P_j \equiv C^\dagger Z_j C,
\label{eq:pauliconj}
\end{equation}
we note that the $Z_j$ mutually commute, implying $P_j$ mutually commute as well. Then, we can use the form of the coherent errors~\eqref{eq:coherent_error} to obtain
\begin{align}
C^\dagger V_{\alpha} C
&= C^\dagger\left(\prod_{j=1}^N e^{-i\frac{\alpha}{2}Z_j}\right)C
= \prod_{j=1}^N e^{-i\frac{\alpha}{2}C^\dagger Z_j C} \nonumber\\
&= \prod_{j=1}^N e^{-i\frac{\alpha}{2}P_j}
= \prod_{j=1}^N \left[c_\alpha\,I- i s_\alpha\,P_j \right],
\end{align}
where $c_\alpha\equiv \cos(\alpha/2)$ and $s_\alpha\equiv \sin(\alpha/2)$.
Expanding the product and grouping terms by Hamming weight yields
\begin{equation}
\prod_{j=1}^N \bigl[c_\alpha\,I- i s_\alpha\,P_j\bigr]
=
\sum_{\ell=0}^N (-i s_\alpha)^\ell c_\alpha^{N-\ell}
\sum_{|A|=\ell} P_A,
\label{eq:expa}
\end{equation}
where $A\subseteq\{1,\dots,N\}$ is a subset of sites, $P_A\equiv \prod_{j\in A}P_j$, and $P_{\emptyset}=I$.
For the case of interest $|\psi\rangle_L=|0\rangle^{\otimes k}$, and thus $|\Psi_0\rangle = |0\rangle^{N}$, the post-measurement logical state after observing syndrome $\mathbf{s}$ reads
\begin{equation}
\ket{\psi_{U}(\mathbf{s})}_L
=
\frac{1}{\sqrt{p_U(\mathbf{s})}}
\sum_{\ell=0}^N (-i s_\alpha)^\ell c_\alpha^{N-\ell}
\sum_{|A|=\ell} \bra{\mathbf{s}}_S P_A \ket{0}^{\otimes N}.
\end{equation}

The operator $P_A$, being a product of Pauli strings, is itself an $N$-qubit Pauli operator. We parameterize it as
\begin{equation}
P_A = \omega_A\prod_{j=1}^N X_j^{\,n_j^A} Z_j^{\,m_j^A},
\end{equation}
where $\omega_A\in\{\pm1\}$ and $n_j^A,m_j^A\in\{0,1\}$ define bitstrings $\mathbf{n}^A=(n_j^A)\in\mathbb{Z}_2^N$ and $\mathbf{m}^A=(m_j^A)\in\mathbb{Z}_2^N$.
The action of $P_A$ on the state $\ket{0}^{\otimes N}$ reduces to bit flips due to the $X_j$ operators, so that
\begin{equation}
P_A\ket{0}^{\otimes N}
= \omega_A\prod_j X_j^{\,n_j^A}\ket{0}^{\otimes N}
\equiv \omega_A\ket{\mathbf{n}^A}.
\end{equation}
Projecting onto a syndrome state $\ket{\mathbf{s}}$ selects the components consistent with $\mathbf{s}$. Writing $\mathbf{n}^A=(\mathbf{x}^A,\mathbf{y}^A)$, with $\mathbf{x}^A\in\mathbb{Z}_2^{k}$ ($\mathbf{y}^A\in\mathbb{Z}_2^{N-k}$) the restriction to the logical (syndrome) subsystem, we have
\begin{equation}
\langle\mathbf{s}|_S\cdot |\mathbf{n}^A\rangle
=\delta_{\mathbf{s},\,\mathbf{y}^A}\,\ket{\mathbf{x}^A}_L,
\label{eq:oversn}
\end{equation}
which yields
\begin{equation}
\ket{\psi_{U}(\mathbf{s})}_L
=
\frac{1}{\sqrt{p_U(\mathbf{s})}}
\sum_{\ell=0}^N (-i s_\alpha)^\ell c_\alpha^{N-\ell}
\sum_{|A|=\ell}\delta_{\mathbf{s},\,\mathbf{y}^A}\,\omega_A\ket{\mathbf{x}^A}_L.
\label{eq:PauliProp}
\end{equation}
Eq.~\eqref{eq:PauliProp} is the starting point of our numerical approach for Clifford encoding--decoding circuits: for each $\ell$ we sum over subsets $A$ with $|A|=\ell$, collect the contributions to each logical computational-basis state $\ket{\mathbf{x}^A}$ consistent with the observed syndrome $\mathbf{s}$, and then sum over $\ell$.

Numerically, evaluating Eq.~\eqref{eq:PauliProp} requires manipulating only the logical state vector $\ket{\psi_U(\mathbf{s})}_L$ in a $2^k$-dimensional space. This enables us to access systems of up to $N=44$ qubits at code rate $r=k/N=1/2$.

Conceptually, Eq.~\eqref{eq:PauliProp} reveals a nontrivial structure of \textit{syndrome classes} for Clifford encoders. We say that a syndrome $\mathbf{s}$ has class $\ell_{\mathbf{s}}$ if the smallest subset size $|A|$ contributing nontrivially to the sum in Eq.~\eqref{eq:PauliProp} is $\ell_{\mathbf{s}}$. In particular, for the input $\ket{0}^{\otimes N}$ the trivial syndrome $\mathbf{0}\equiv(0,\ldots,0)\in\mathbb{Z}_2^{N-k}$ has class $\ell_{\mathbf{0}}=0$, since the $\ell=0$ term contributes $c_\alpha^{N}\ket{0}^{\otimes k}$ to the final logical state $\ket{\psi_U(\mathbf{0})}_L$. In contrast, any nontrivial syndrome $\mathbf{s}\neq \mathbf{0}$ satisfies $\ell_{\mathbf{s}}\ge 1$, because at least one non-identity Pauli factor is required to produce a nonvanishing contribution $\delta_{\mathbf{s},\,\mathbf{y}^A}\ket{\mathbf{x}^A}$.

\section{Forced Measurements and Encoding-Decoding Transitions}
\label{sec:forced}
We begin by studying the ``forced'' measurement setting, in which we post-select the syndrome readout onto a prescribed bitstring $\mathbf{s}$. We first focus on the trivial syndrome $\mathbf{s}=\mathbf{0}$. In quantum error-correction language, this is the \textit{success} branch of the decoding procedure: the decoder detects no syndrome (equivalently, it infers that no error has occurred, or that any error has been perfectly corrected). 
Later, we turn to nontrivial outcomes $\mathbf{s}\neq\mathbf{0}$ and characterize the resulting logical states $\ket{\psi_{U}(\mathbf{s})}_L$.

\subsection{Fidelity}
\label{subsec:fidelity}
The central diagnostic of the error-resilience transition is the logical fidelity $F_U(\mathbf{s})$ in Eq.~\eqref{eq:Fs}. In the error-resilient phase, decoding successfully inverts the scrambling and error layers, yielding $F_U\to 1$ in the thermodynamic limit. In the error-vulnerable phase, logical information is effectively erased and the fidelity drops to its random-state value, $F_U\to 2^{-k}$. The corresponding behavior of ensemble-averaged fidelities for Haar encoders was obtained analytically in Ref.~\cite{turkeshi2024error}. Here we go beyond averages and analyze both the circuit-averaged fidelity and its sample-to-sample fluctuations, establishing self-averaging and providing a complete analytical description of the error-resilience transition for both Haar and Clifford encoding-decoding circuits.

\subsubsection{Annealed fidelity for Haar-Random Encoders}
\label{subsubsec:forced_haar}

Starting from the definition of the fidelity in Eq.~\eqref{eq:Fs} and using the post-selected logical state [cf.
Eq.~\eqref{eq:stateS1}], we can express the fidelity for a fixed encoder $U$ and syndrome $\mathbf{s}$ as the ratio
\begin{equation}
F_{U}(\mathbf{s})=\frac{m_{F,U}(\mathbf{s})}{p_{U}(\mathbf{s})}.
\label{eq:F_ratio}
\end{equation}
Here, the denominator is the Born probability of obtaining the outcome $\mathbf{s}$,
\begin{equation}
\begin{split}
p_{U}(\mathbf{s})
&=\bigl\|\bra{\mathbf{s}}_S\,U^\dagger V_\alpha U\,\ket{\Psi_0}\bigr\|^2
\\&=
\bra{\Psi_0}\,U^\dagger V_\alpha^\dagger U\,\Pi_{I,\mathbf{s}}\,U^\dagger V_\alpha U\,\ket{\Psi_0},
\label{eq:pF_def}
\end{split}
\end{equation}
while the numerator is the overlap with the target logical state $\ket{\psi}_L$ within the corresponding post-selected branch,
\begin{equation}
\begin{split}
m_{F,U}(\mathbf{s})
&=
\bigl|(\bra{\psi|_L\otimes\langle \mathbf{s}}_S)\,U^\dagger V_\alpha U\,\ket{\Psi_0}\bigr|^2
\\
&=
\bra{\Psi_0}\,U^\dagger V_\alpha^\dagger U\,\Pi_{\psi,\mathbf{s}}\,U^\dagger V_\alpha U\,\ket{\Psi_0},
\label{eq:mF_def}
\end{split}
\end{equation}
with $\Pi_{\psi,\mathbf{s}} =(\ket{\psi}\!\bra{\psi})_L\otimes (\ket{\mathbf{s}}\!\bra{\mathbf{s}})_S$.

Following the general discussion in \textbf{Sec.}~\ref{sec:methods}, we define the \emph{quenched fidelity} by averaging the full ratio in Eq.~\eqref{eq:F_ratio} over the ensemble of encoders,
\begin{equation}
\overline{F}(\mathbf{s})
=
\mathbb{E}_{U\sim\mathcal{E}}\!\left[\frac{m_{F,U}(\mathbf{s})}{p_{U}(\mathbf{s})}\right].
\label{eq:F_quenched}
\end{equation}
In contrast, the \emph{annealed fidelity} is defined by 
\begin{equation}
\tilde{F}(\mathbf{s})
\equiv
\frac{\mathbb{E}_{U\sim\mathcal{E}}\!\left[m_{F,U}(\mathbf{s})\right]}
{\mathbb{E}_{U\sim\mathcal{E}}\!\left[p_{U}(\mathbf{s})\right]}.
\label{eq:F_annealed}
\end{equation}
As anticipated in \textbf{Sec.}~\ref{sec:methods}, in general $\overline{F}(\mathbf{s})\neq \tilde{F}(\mathbf{s})$, since averaging does not commute with taking ratios. 
We now specialize to the forced branch $\mathbf{s}=\mathbf{0}$, in which case, the quench and annealed value coincide in the thermodynamic limit.

Considering two copies $\mathcal{H}^{\otimes 2}$ of the system Hilbert space, we introduce the two-replica operator
\begin{equation}
A_U^{(2)} \equiv (U^\dagger)^{\otimes 2}\,(V_\alpha\otimes V_\alpha^\dagger)\,U^{\otimes 2}.
\label{eq:A2_def}
\end{equation}
This operator linearizes the squared amplitudes in Eqs.~\eqref{eq:pF_def} and~\eqref{eq:mF_def} as traces over the doubled Hilbert space. 
Without loss of generality, we can fix through Haar invariance the initial logical state to $|\psi\rangle_L = \ket{\mathbf{0}}_L$. 
Let us introduce the two-replica operator \begin{equation}
   \mathcal{B}(O)\equiv \bigl(|\mathbf{0}\rangle\langle\mathbf{0}|_L\otimes |\mathbf{0}\rangle\langle\mathbf{0}|_S\bigr)\otimes \bigl(O_L\otimes |\mathbf{0}\rangle\langle\mathbf{0}|_S\bigr)
   \label{eq:BA}
\end{equation}
where $O$ is an arbitrary operator acting on the logical space. 
Then, the replica boundary conditions for the fidelity are
\begin{equation}
\mathcal{B}_{\mathrm{num}}^{F}  \equiv  T_{(12)} \mathcal{B}(|\mathbf{0}\rangle\langle\mathbf{0}|),
\qquad
\mathcal{B}_{\mathrm{den}}^{F} \equiv  T_{(12)} \mathcal{B}(I)\;,
\label{eq:fidB}
\end{equation}
where $T_{(12)}$ swaps the two replicas.
Expressing the numerator and denominator in Eq.~\eqref{eq:F_annealed} as traces of $A_U^{(2)}$ with the boundary operators, cf. Eq.~\eqref{eq:fidB}, and using linearity to interchange $\mathbb{E}_{U}$ with the trace, we obtain the replica representation of the annealed fidelity,
\begin{equation}
\tilde{F}(\mathbf{0})
=
\frac{\Tr\!\left(\mathcal{B}_{\mathrm{num}}^{F}\, \mathbb{E}_{U\sim \mathcal{E}} \left[ A_U^{(2)} \right]\right)}
{\Tr\!\left(\mathcal{B}_{\mathrm{den}}^{F}\,  \mathbb{E}_{U\sim \mathcal{E}} \left[ A_U^{(2)} \right]\right)}.
\label{eq:F_replica_ratio}
\end{equation}
This representation is convenient because computing $\tilde{F}(\mathbf{0})$ reduces to averaging the operator $A_U^{(2)}$, which can be carried out using Weingarten calculus~\cite{Collins2006,Roberts17}, followed by contraction with the boundary operators $\mathcal{B}_{\mathrm{num}}^{F}, \,\mathcal{B}_{\mathrm{den}}^{F}$ and taking the trace.

For the unitary-group ensemble $\mathcal{E}=\mathcal{U}(2^N)$, Schur--Weyl duality gives the two-copy twirl
\begin{equation}
\mathbb{E}_{U\sim \mathcal{U}(2^N)}\!\left[A_U^{(2)}\right]
=
\sum_{\pi,\sigma\in \mathrm{S}_2}
\mathrm{Wg}_{\pi,\sigma}\,
\Tr\!\left(T_{\pi}\,V_\alpha\otimes V_\alpha^\dagger\right)\,
T_{\sigma},
\label{eq:Afid}
\end{equation}
where $\mathrm{S}_2=\{(),(12)\}$ is the permutation group of two elements, $T_{\sigma}$ denotes the permutation representation in replica space, and the 
coefficients $\mathrm{Wg}_{\pi,\sigma}$ are known as Weingarten functions~\cite{Weingarten78, Collins2006}, see \textbf{App.}~\ref{App:averages} for details.
Using the coherent-error form~\eqref{eq:coherent_error}, and specializing to $N\ge k$, we find
$
\Tr\!\left(T_{()}\,V_\alpha\otimes V_\alpha^\dagger\right)
=\bigl|\Tr(V_\alpha)\bigr|^2
=\bigl[2\cos(\alpha/2)\bigr]^{2N}$, and 
$
\Tr\!\left(T_{(12)}\,V_\alpha\otimes V_\alpha^\dagger\right)
=\Tr(V_\alpha V_\alpha^\dagger)
=2^N$.
Moreover, for $d=2^N$ the Weingarten coefficients for $\mathrm{S}_2$ are
$
\mathrm{Wg}_{(),()}=\mathrm{Wg}_{(12),(12)}=(d^2-1)^{-1}$, and
$\mathrm{Wg}_{(),(12)}=\mathrm{Wg}_{(12),()}=-[d(d^2-1)]^{-1}$.
To complete the evaluation of $\tilde{F}(\mathbf{0})$, it remains to compute the contractions of the boundary operators with $T_\sigma$:
$
\Tr\!\left(\mathcal{B}_{\mathrm{num}}^{F}\,T_{()}\right)
=\Tr\!\left(\mathcal{B}_{\mathrm{num}}^{F}\,T_{(12)}\right)=1$,
and $
\Tr\!\left(\mathcal{B}_{\mathrm{den}}^{F}\,T_{()}\right)=1$,
$\Tr\!\left(\mathcal{B}_{\mathrm{den}}^{F}\,T_{(12)}\right)=2^k$.
We note that these formulas can be evaluated analytically since both permutation representation factorizes over qubits, i.e., $T_\sigma = t_\sigma^{\otimes N}$ (where $t_\sigma$ acts on  a single qubit of the first and the second replica), and also the boundary operators $ \mathcal{B}_{\mathrm{num}}^{F}$, $ \mathcal{B}_{\mathrm{den}}^{F} $, are tensor products of $N$ operators, each acting in the single-qubit replica space.
The above identities yield
\begin{equation}
\label{eq:resN}
\begin{split}
   &\mathbb{E}_{U}[m_{F,U}(\mathbf{0})]
    =\frac{2^N \cos^{2N}(\alpha/2)+1}{2^N+1}
\end{split}
\end{equation}
 for the numerator, while for the denominator we have
\begin{equation}
\label{eq:resD}
\begin{split}
     &\mathbb{E}_{U}[p_{U}(\mathbf{0})]
    =\frac{2^N (2^N-2^k)\cos^{2N}(\alpha/2)+2^{N+k}-1}{2^{2N}-1}
\end{split}
\end{equation}

\begin{figure*}
    \centering
    \includegraphics[width=1\linewidth]{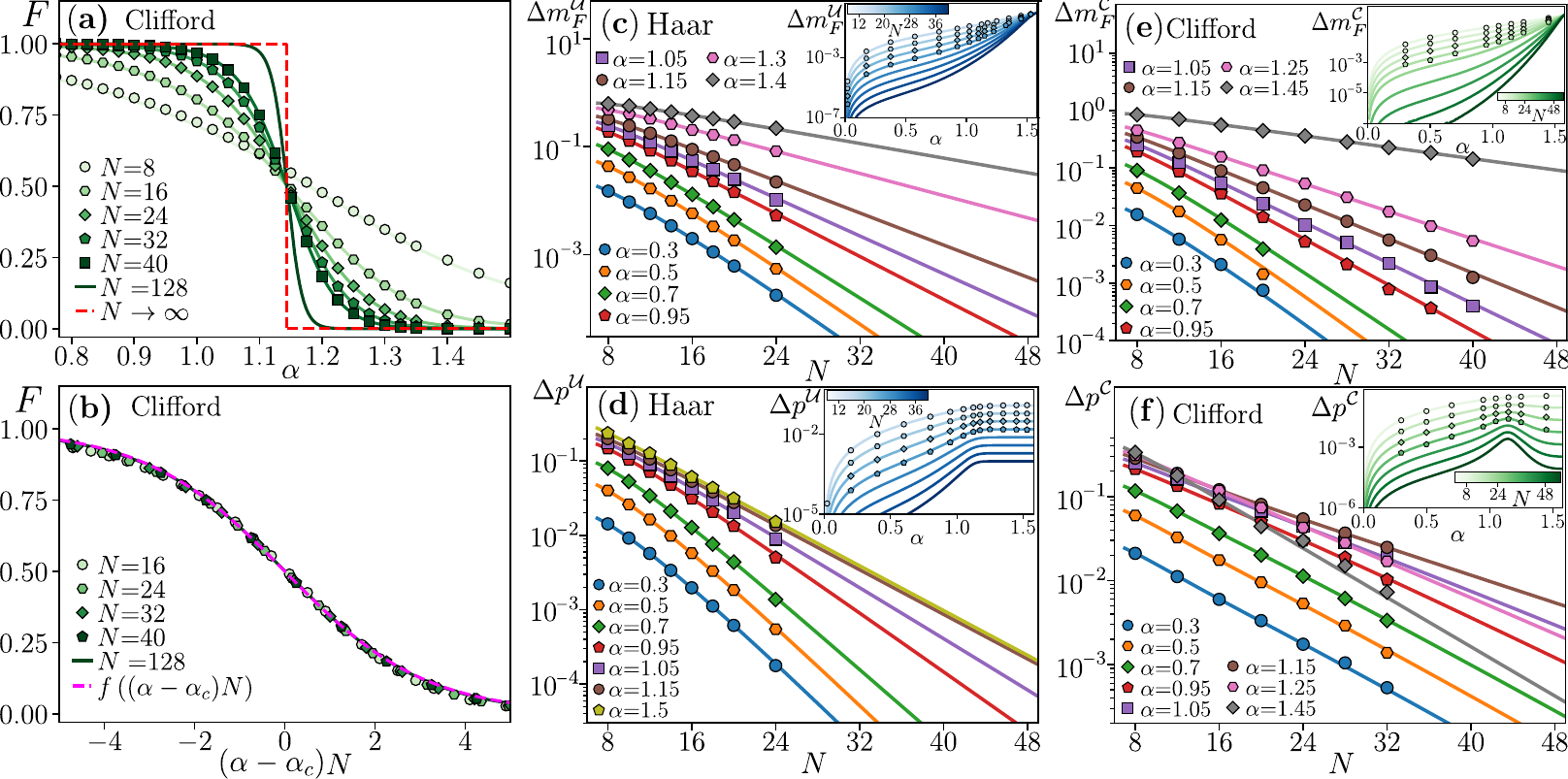}
    \caption{
Error-resilience phase transition in encoding-decoding circuits: fidelity and self-averaging. (\textbf{a}) Fidelity $F$ as function of error strength $\alpha$ for system size $N$. (\textbf{b}) Finite-size scaling collapses of fidelity. Panels (\textbf{c}) and (\textbf{d}) show numerator and denominator fluctuations for Haar encoders at fixed error strenght $\alpha$, as a function of system size $N$. (\textbf{e}) and (\textbf{f}) show numerator and denominator fluctuations for the Clifford encoders. The insets in (\textbf{c}-\textbf{f}) show the corresponding data but plotted as function of $\alpha$ for various system sizes $N$. In all panels the solid lines show analytical expressions obtained with the replica method, while markers denote the numerical results averaged over more than 1000 circuit realizations.
    }
    \label{fig:fide}
\end{figure*}

Eqs.~\eqref{eq:resN} and~\eqref{eq:resD} provide an exact closed-form expression for the annealed fidelity $\tilde{F}(\mathbf{0})$ at arbitrary system size $N$ and logical-qubit number $k$. In the scaling limit $k=rN$ with $N\gg 1$, it is convenient to define the critical error strength $\alpha_c$ by
$
\cos\!\left(\alpha_c/2\right)=2^{\frac{r-1}{2}} .
$
To leading order at large $N$, one then finds
\begin{equation}
\tilde{F}(\mathbf{0})
=
\left[
1+
\left(
\frac{\cos(\alpha/2)}{\cos(\alpha_c/2)}
\right)^{-2N}
\right]^{-1}.
\end{equation}
Expanding near $\alpha=\alpha_c$ yields the scaling form
\begin{equation}
\tilde{F}(\mathbf{0})
=
\left[
1+\exp\,\!\Bigl(\tan\!\bigl(\tfrac{\alpha_c}{2}\bigr)\,N(\alpha-\alpha_c)\Bigr)
\right]^{-1}.
\label{eq:uniF}
\end{equation}
Equation~\eqref{eq:uniF} shows that $\tilde{F}(\mathbf{0})\to 1$ for $\alpha<\alpha_c$ and $\tilde{F}(\mathbf{0})\to 0$ for $\alpha>\alpha_c$, capturing the error-resilience transition. Moreover, at large $N$ the fidelity depends only on the scaling variable $N(\alpha-\alpha_c)$, implying a correlation-length exponent $\nu=1$.

\subsubsection{Annealed fidelity for Clifford-Random Encoders}
\label{subsubsec:forced_cliff}
We now turn to the annealed fidelity $\tilde{F}(\mathbf{0})$ for Clifford encoders. The $N$-qubit Clifford group forms a unitary $2$-design~\cite{Gross07design,Dankert09,zhu2017multiqubitcliffordgroups,webb2016clifford}, meaning that its second moments match those of the Haar measure~\cite{Mele2024introductiontohaar}. Equivalently, in the language of Schur--Weyl duality and commutant structure~\cite{Gross2021,Bittel25commutant}, the two-copy Clifford twirl involves the same permutation algebra as the Haar twirl, and therefore defines the same two-replica twirling channel. In particular, the Clifford average of the two-copy operator $A_U^{(2)}$ coincides with the Haar result,
\begin{equation}
\mathbb{E}_{U\sim \mathcal{C}_N}\!\left[A_U^{(2)}\right]
=
\sum_{\pi,\sigma\in \mathrm{S}_2}
\mathrm{Wg}_{\pi,\sigma}\,
\Tr\!\left(T_{\pi}\,V_\alpha\otimes V_\alpha^\dagger\right)\,
T_{\sigma},
\label{eq:Acli2}
\end{equation}
and hence the annealed fidelity for Clifford encoders is identical to Eqs.~\eqref{eq:resN} and~\eqref{eq:resD}. It follows that $\tilde{F}(\mathbf{0})$ exhibits the same error-resilience transition, with the same critical error strength $\alpha_c$ and finite-size scaling form [cf. Eq.~\eqref{eq:uniF}], implying $\nu=1$.

In Fig.~\ref{fig:fide}(a), at fixed code rate $r=1/2$, we compare the quenched fidelity for the Clifford encoders $\overline{F}(\mathbf{0})$ obtained numerically through Eq.\eqref{eq:PauliProp} (symbols), with the analytical annealed prediction $\tilde{F}(\mathbf{0})$ from Eqs.~\eqref{eq:resN} and~\eqref{eq:resD} (solid lines). Already for $N=8$, $\overline{F}(\mathbf{0})$ and $\tilde{F}(\mathbf{0})$ are in close agreement, and their discrepancy decreases with increasing system size. As a function of the coherent error strength $\alpha$, the fidelity sharpens into an increasingly step-like profile, signaling the error-resilience transition. In Fig.~\ref{fig:fide}(b), rescaling the control parameter as $(\alpha-\alpha_c)N^{1/\nu}$ with $\nu=1$ collapses the data for different $N$ onto the universal scaling curve, cf. Eq.~\eqref{eq:uniF}.

Although the annealed prediction $\tilde{F}(\mathbf{0})$ is identical for Haar and Clifford encoders (by the $2$-design property), the finite-$N$ quenched estimates $\overline{F}(\mathbf{0})$ show small ensemble-dependent deviations at the smallest sizes. Numerically, these differences decrease with $N$~\footnote{See also the data for the Haar case in Ref.~\cite{turkeshi2024error}.}.
In the next section, we show that $\tilde{F}(\mathbf{0})$ and $\overline{F}(\mathbf{0})$ coincide in the thermodynamic limit, for both Haar and Clifford encoders.

\subsubsection{Self-averaging of fidelity}
\label{subsubsec:self}

To explain the observed proximity, and eventual equality in the thermodynamic limit, of the annealed and quenched fidelities, we use the fact that the fidelity is a ratio of two random variables, Eq.~\eqref{eq:F_ratio}. 
The quenched prescription averages the ratio $m_{F,U}(\mathbf{s})/p_{U}(\mathbf{s})$ over the ensemble, whereas the annealed prescription forms the ratio of the separate ensemble averages of $m_{F,U}(\mathbf{s})$ and $p_{U}(\mathbf{s})$. We will show that both $m_{F,U}(\mathbf{0})$ and $p_{U}(\mathbf{0})$ are self averaging, meaning that their relative fluctuations around the mean vanish as $N\to\infty$. In this regime the ratio $F_U$ concentrates sharply, and the two averaging procedures coincide.

For notational simplicity, throughout this section we suppress the explicit dependence on the post selected trivial syndrome and write $F$, $m_{F,U}$, and $p_U$ in place of $F(\mathbf{0})$, $m_{F,U}(\mathbf{0})$, and $p_{U}(\mathbf{0})$.

We quantify realization to realization fluctuations of $m_{F,U}$ and $p_{U}$ for an ensemble $\mathcal{E}$ of encoders through the relative variances
\begin{equation}
\bigl[\Delta m^{\mathcal{E}}_F\bigr]^2
=
\frac{\mathbb{E}_{U\sim \mathcal{E}}\!\left[m_{F,U}^2\right]
-\left(\mathbb{E}_{U\sim \mathcal{E}}\!\left[m_{F,U}\right]\right)^2}
{\left(\mathbb{E}_{U\sim \mathcal{E}}\!\left[m_{F,U}\right]\right)^2},
\label{eq:fluct1}
\end{equation}
and
\begin{equation}
\bigl[\Delta p^{\mathcal{E}}\bigr]^2
=
\frac{\mathbb{E}_{U\sim \mathcal{E}}\!\left[p_{U}^2\right]
-\left(\mathbb{E}_{U\sim \mathcal{E}}\!\left[p_{U}\right]\right)^2}
{\left(\mathbb{E}_{U\sim \mathcal{E}}\!\left[p_{U}\right]\right)^2}.
\label{eq:fluct2}
\end{equation}
Computing $\bigl[\Delta m^{\mathcal{E}}_F\bigr]^2$ and $\bigl[\Delta p^{\mathcal{E}}\bigr]^2$ requires (i) the first moments
$\mathbb{E}_{U\sim \mathcal{E}}[m_{F,U}]$ and $\mathbb{E}_{U\sim \mathcal{E}}[p_{U}]$, which were obtained in the preceding section, and (ii) the second moments
$\mathbb{E}_{U\sim \mathcal{E}}[m_{F,U}^2]$ and $\mathbb{E}_{U\sim \mathcal{E}}[p_{U}^2]$.
The latter can be written as traces in the replica space $\mathcal{H}^{\otimes 4}$ involving the four copy operator
\begin{equation}
A_U^{(4)}
\equiv
(U^\dagger)^{\otimes 4}\,
\bigl(V_\alpha \otimes V_\alpha^\dagger\bigr)^{\otimes 2}\,
U^{\otimes 4},
\label{eq:A4_def}
\end{equation} 
while the suitable boundary replica operators are given as $B^{m_F} =  T_{(12)(34)}\mathcal{B}(|\mathbf{0}\rangle\langle\mathbf{0}|)^{\otimes 2}$ and 
$
\mathcal{B}^{p_F}
=
T_{(12)(34)} \mathcal{B}(I)^{\otimes 2}$, see \textbf{App.}~\ref{app:fideB} for details, 
enabling the analytical calculation of $\mathbb{E}_{U\sim \mathcal{E}}[m_{F,U}^2]$ and $\mathbb{E}_{U\sim \mathcal{E}}[p_{U}^2]$.

For Haar encoders, $\mathcal{E}=\mathcal{U}(2^N)$, the average of Eq.~\eqref{eq:A4_def} has the same structure as Eq.~\eqref{eq:Afid}, with the permutation group $\mathrm{S}_2$ replaced by $\mathrm{S}_4$. Since $|\mathrm{S}_4|=4!=24$, the full expressions (reported in~\cite{OURzenodo}) are lengthy; here we quote the leading behavior at $N\gg1$,
\begin{equation}
\bigl[\Delta m^{\mathcal{U}}_F\bigr]^2
=
\frac{1 + 2^{1+N}\cos^{2N}(\alpha/2)}
{\bigl(1 + 2^{N}\cos^{2N}(\alpha/2)\bigr)^2},
\end{equation}
and
\begin{equation}
\bigl[\Delta p^{\mathcal{U}}\bigr]^2
=
\frac{2^{Nr} + 2^{1+N}\cos^{2N}(\alpha/2)\bigl(1+\cos^{2N}(\alpha/2)\bigr)}
{\bigl(2^{Nr} + 2^{N}\cos^{2N}(\alpha/2)\bigr)^2}.
\end{equation}

The Clifford group does not form a unitary 4 design. This implies that, from the $n\ge 4$, the $n$-th Clifford twirl involves a set of reduced Pauli monomials $\mathfrak{P}_n = \{ \Omega \}$ which, in addition to permutations, contains extra ``intrinsic'' Clifford commutant operators~\cite{Gross2021,Bittel25commutant,Turkeshi_2024_2}, as summarized in \textbf{App.}~\ref{app:cliffINT}.
The cardinality of the set of reduced Pauli monomials  is
\begin{equation}
|\mathfrak{P}_n| = \prod_{m=0}^{n-2} \left(2^m + 1\right),
\label{eq:Sigma_k_cardinality}
\end{equation}
which is strictly larger than the permutation group size $|\mathrm{S}_n|=n!$ for any $n\geq 4$. 
The average of $A_U^{(4)}$ over the Clifford group $\mathcal{C}_N$ reads
\begin{equation}
\mathbb{E}_{U\sim \mathcal{C}_N}\!\left[A_U^{(4)}\right]
=
\sum_{\Omega,\Omega' \in \mathfrak{P}_4}
\mathrm{Wg}^{\mathcal{C}_N}_{\Omega,\Omega'}\,
\Tr\!\left( \Omega \,\bigl(V_\alpha \otimes V_\alpha^\dagger\bigr)^{\otimes 2}\right)\,
{\Omega'},
\label{eq:Acli}
\end{equation}
where the coefficients $\mathrm{Wg}^{\mathcal{C}_N}_{\Omega,\Omega'}$ form a Clifford Weingarten matrix which is obtained as the inverse of Clifford Gram matrix $G$ with elements $G^{\mathcal{C}_N}_{\Omega,\Omega'}=\Tr[\Omega ({\Omega'})^\dagger]$ for $\Omega, \Omega'\in \mathfrak{P}_4$.
Together with the permutation operators $T_\sigma$ with $\sigma\in \mathrm{S}_4$, $\mathfrak{P}_4$ induces additional 6 operators of the form $T_\tau\Omega_{(1,1,1,1)}$ for $\tau \in \mathrm{S}_3\otimes \mathrm{S}_1$, where $\Omega_{(1,1,1,1)}\equiv \sum_{P\in \mathcal{P}_N}P^{\otimes 4}/2^N$~\cite{roth2018recoveringquantumgates, Leone2021quantumchaosis,Turkeshi_2023,Haug_2023_1,Lami_2023_2,Lami_2024,tarabunga2023manybody,tarabunga2024nonstabilizerness}.

Noticing that $\Omega_{(1,1,1,1)}=\left[\sum_{P\in\mathcal{P}_1} \left(P^{\otimes4}/2\right)\right]^{\otimes N}$, similarly to the permutation representation $T_\sigma$, admits a tensor product form over the $N$ qubits, we can now calculate the analytical average for the fidelity numerator and denominator fluctuations in the Clifford encoder case exactly.
The resulting expressions (reported in full in~\cite{OURzenodo}) differ from those obtained in the Haar case.
The leading contribution reads
\begin{equation}
\bigl[\Delta m^{\mathcal{C}}_F\bigr]^2
=
\frac{3 + 2^{1+N}\cos^{2N}(\alpha/2) + 2^{-N}\bigl(3+\cos(2\alpha)\bigr)^{N}}
{\bigl(1 + 2^{N}\cos^{2N}(\alpha/2)\bigr)^2},
\end{equation}
and
\begin{equation}
\bigl[\Delta p^{\mathcal{C}}\bigr]^2
=
\frac{
2^{N(r-1)}
\left(
2^{N}
-4^{N}\cos^{4N}(\alpha/2)
+\bigl(3+\cos(2\alpha)\bigr)^{N}
\right)
}{
\left(2^{Nr}+2^{N}\cos^{2N}(\alpha/2)\right)^{2}
}.
\end{equation}

The relative fluctuations $\Delta m^{\mathcal{E}}_F$ and $\Delta p^{\mathcal{E}}$, shown for $r=1/2$ in Fig.~\ref{fig:fide}(c)--(f), decay exponentially with $N$ for all $\alpha$ (with an $\alpha$-dependent rate). This exponential suppression demonstrates that both $m_{F,U}$ and $p_{U}$ are self averaging. Accordingly, we write
\[
m_{F,U}=\mathbb{E}_U[m_{F,U}]\,(1+\delta_m),
\qquad
p_{U}=\mathbb{E}_U[p_{U}]\,(1+\delta_p),
\]
where typically $\delta_m\sim \Delta m_F^{\mathcal{E}}$ and $\delta_p\sim \Delta p^{\mathcal{E}}$. Expanding the ratio gives
\begin{equation}
\frac{m_{F,U}}{p_{U}}
=
\frac{\mathbb{E}_U[m_{F,U}]}{\mathbb{E}_U[p_{U}]}\,
\frac{1+\delta_m}{1+\delta_p}
=
\tilde F\left(1+\delta_m-\delta_p+O(\delta_m^2,\delta_p^2)\right).
\label{eq:Fselfav}
\end{equation}
Taking the ensemble average of Eq.~\eqref{eq:Fselfav}, the left-hand side yields the quenched fidelity $\overline{F}$, while the prefactor on the right-hand side is the annealed fidelity $\tilde F$. Since $\delta_m$ and $\delta_p$ vanish exponentially with $N$, the correction factor approaches unity as $N\to\infty$, implying $\overline{F}\to \tilde F$ in the thermodynamic limit. Thus, both averaging prescriptions define the same asymptotic order parameter for the error-resilience transition.

The error-resilience transition, across which $\overline{F}$ jumps from unity in the error-resilient phase ($\alpha<\alpha_c$) to zero in the error-vulnerable phase ($\alpha>\alpha_c$), is accompanied by a qualitative reorganization of the post-selected logical state $\ket{\psi(\mathbf{0})}_L$. For an initial logical basis state $\ket{\mathbf{0}_L}$, the output $\ket{\psi(\mathbf{0})}_L$ remains unentangled and sharply localized in the computational basis for $\alpha<\alpha_c$, whereas for $\alpha>\alpha_c$ it develops multifractal structure. This is captured by participation entropies~\cite{Liu25,sierant2022universalbehaviorbeyond,Turkeshi_2024_hiler,lami2025anticoncentration,mace2019multifractal,luitz2014universalbehaviorbeyond}, which probe higher moments and therefore depend on averages involving four (and more) replicas. As a result, participation-entropic diagnostics differ between Haar and Clifford ensembles: the Haar case is characterized in Ref.~\cite{turkeshi2024error}, while the Clifford-encoder results are presented in \textbf{App.}~\ref{app:parti}. 
The fact that the critical points extracted from the fidelity and from the participation entropies coincide indicates that a single underlying mechanism, the error-resilience threshold at $\alpha_c$, controls the behavior of the logical state in the forced measurement case $\mathbf{s}=\mathbf{0}$.

\subsection{Magic resources phase transition}
\label{subsec:magica}

\begin{figure*}
    \centering
    \includegraphics[width=1\linewidth]{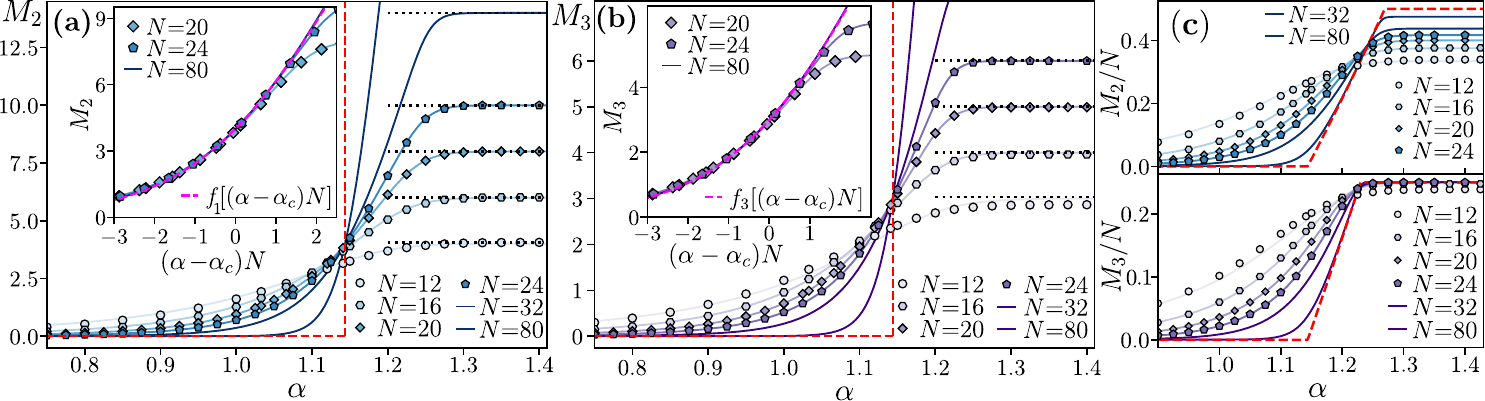}
    \caption{
Phase transition in magic resources for Haar-random encoders with the postselected syndrome $\mathbf{s}=\mathbf{0}$. (a) The SRE $M_2$ as a function of the error strength $\alpha$ for various system sizes $N$ and a fixed code rate $r=k/N=1/2$. Solid lines represent the analytic formula~\eqref{eq:SRE2good}, while symbols denote the quenched averages $\overline{M}_2$ computed via state-vector simulations averaged over 1000 circuit realizations. The inset shows the collapse of $M_2$ onto the universal curve~\eqref{eq:M2_haar_scaling} when plotted against the rescaled error strength $(\alpha-\alpha_c)N^{1/\nu}$, with $\nu=1$. (b) Analogous results for the SRE $M_3$. Panels (c) and (d) show the SRE densities $M_q/N$ for $q=2$ and $q=3$, respectively, with the red dashed lines indicating the behavior in the thermodynamic limit ($N\to\infty$).
}
    \label{fig:SRE2H}
\end{figure*}

In this section, we turn to analysis of magic state resources of the post-selected logical state $\ket{\psi_U(\mathbf{0})}_L$, quantified by the SRE~\eqref{eq:sreDEF}.
We start by showing how the stabilizer entropy fits into the same replica averaging framework as the fidelity. 
As before, we focus on the forced branch $\mathbf{s}=\mathbf{0}$ and the initial state $\ket{\Psi_0}=\ket{0}^{\otimes N}$.

The \textit{quenched SRE} is obtained by averaging $M_q(\rho_{L,U})$, Eq.~\eqref{eq:sreDEF}, over the encoder ensemble,
\begin{equation}
\overline{M}_q(\mathbf{0})
\equiv
\mathbb{E}_{U\sim\mathcal{E}}\!\left[M_2\!\left(\rho_{L,U}\right)\right].
\label{eq:M2_quenched}
\end{equation}

Using the form of the normalized logical state~\eqref{eq:rhoL} in the definition of the SRE, we find that the SRE is 
\begin{equation}
    M_q(\rho_{L,U}) = \frac{1}{1-q} \log_2 \left[ \frac{m^{(q)}_{M,U}(\mathbf{s})}{p^{2q}_U(\mathbf{s})}\right],
    \label{eq:sreAN1}
\end{equation}
where $p_U(\mathbf{s})$ is given by Eq.~\eqref{eq:pF_def}, and the numerator reads
\begin{equation}
    m^{(q)}_{M,U}(\mathbf{s}) =\frac{1}{2^k} \sum_{P \in \mathcal{P}_k} \bra{\Psi_0}\,U^\dagger V_\alpha^\dagger U\,\Pi_{P,\mathbf{s}}\,U^\dagger V_\alpha U\,\ket{\Psi_0}^{2q},
    \label{eq:sreAN2}
\end{equation}
where $\Pi_{P,\mathbf{s}} =(P)_L\otimes (\ket{\mathbf{s}}\!\bra{\mathbf{s}})_S$.
This motivates us to define the \textit{annealed SRE} as 
\begin{equation}
    \tilde{M}_q(\mathbf{s}) = \frac{1}{1-q} \log_2 \left[ \frac{
    \mathbb{E}_{U\sim\mathcal{E}} \bigl[m^{(q)}_{M,U}(\mathbf{s}) \bigr]}{ \mathbb{E}_{U\sim\mathcal{E}} \bigl[p^{2q}_U(\mathbf{s}) \bigr]} \right].
    \label{eq:sreAN3}
\end{equation}

To analytically evaluate the annealed SRE, we introduce the $4q$-copy Hilbert space
\(
\mathcal{H}^{\otimes (4q)}
\) 
and consider the operator
\begin{equation}
A_U^{(4q)}
\equiv
(U^\dagger)^{\otimes (4q)}\,
\bigl(V_\alpha\otimes V_\alpha^\dagger\bigr)^{\otimes (2q)}\,
U^{\otimes (4q)}.
\label{eq:A8_def}
\end{equation}
As in the fidelity case, the role of $A_U^{(4q)}$ is to linearize the higher powers of circuit amplitudes as traces over the replicated space. 
The remaining ingredient is the appropriate boundary condition implementing the numerator and denominator in Eq.~\eqref{eq:sreAN3}.
With simple algebraic manipulations, detailed in \textbf{App.}~\ref{app:bound}, one finds that the $2q$-th power of $p_U(\mathbf{s})$ and the numerator $m^{(q)}_{M,U}(\mathbf{s})$ can be implemented by the compact boundary operators
\begin{equation}
\begin{split}
\mathcal{B}^{(M_q)}_{\mathrm{num}}&\equiv T_{\sigma_{4q}} \!
\sum_{P\in{\mathcal{P}}_k}\frac{\bigl[\mathcal{B}(P)\bigr]^{\otimes (2q)}}{2^k},
\\
\mathcal{B}^{(M_q)}_{\mathrm{den}}&\equiv T_{\sigma_{4q} } 
\bigl[\mathcal{B}(I)\bigr]^{\otimes (2q)},
\label{eq:B8_def}
\end{split}
\end{equation}
with two-replica operator $\mathcal{B}$ given by Eq.~\eqref{eq:BA}, and $T_{\sigma_{4q}}$ denoting the representation of the pairing permutation $\sigma_{4q}=(12)(34)\ldots(4q-1\ 4q)$.
Combining Eqs.~\eqref{eq:A8_def}--\eqref{eq:B8_def} then yields the replica representation
\begin{equation}
\tilde M_q(\mathbf{0})
= - \log_2 \left( 
\frac{
\Tr\!\left(
\mathcal{B}^{(M_q)}_{\mathrm{num}}\;
\mathbb{E}_{U\sim\mathcal{E}}\!\left[A_U^{(4q)}\right]
\right) 
}{
\Tr\!\left(
\mathcal{B}^{(M_q)}_{\mathrm{den}}\;
\mathbb{E}_{U\sim\mathcal{E}}\!\left[A_U^{(4q)}\right]
\right) 
}\right),
\label{eq:m2_annealed_replica}
\end{equation}
analogous to the annealed fidelity, Eq.~\eqref{eq:F_replica_ratio}. The computation of $\tilde M_q(\mathbf{0})$ therefore reduces to evaluating the $4q$-copy twirl $\mathbb{E}_{U\sim\mathcal{E}}[A_U^{(4q)}]$ via Schur--Weyl duality and Weingarten calculus and contracting it with the fixed boundary operators~\eqref{eq:B8_def}.
In the following we now present the explicit computation for Haar encoders $(\mathcal{E}=\mathcal{U}(2^N))$ and Clifford encoders $(\mathcal{E}=\mathcal{C}_N)$.

\subsubsection{Stabilizer Renyi entropy for Haar encoders}
\label{eq:forzaitalia}

For the unitary-group ensemble $\mathcal{E}=\mathcal{U}(2^N)$, Schur--Weyl duality gives the $4q$-copy twirl in the same form as Eq.~\eqref{eq:Afid}, with $\mathrm{S}_2$ replaced by $\mathrm{S}_{4q}$,
\begin{equation}
\mathbb{E}_{U\sim \mathcal{U}(2^N)}\!\left[A_U^{(4q)}\right]
=\! \!
\sum_{\pi,\sigma\in \mathrm{S}_{4q}} \! \!
\mathrm{Wg}_{\pi,\sigma}
\Tr\!\left(T_{\pi}\,(V_\alpha\otimes V_\alpha^\dagger)^{\otimes (2q)}\right)\,
T_{\sigma}.
\label{eq:A8_twirled}
\end{equation}
In principle, the double sum involves $|\mathrm{S}_{4q}|^2=[(4q)!]^2$ terms and, already for $q=2$ is therefore unwieldy. Instead, we extract the leading behavior in the large-$N$ limit using a \textit{diagonal approximation} of the $4q$-copy Weingarten matrix $\mathrm{Wg}_{\pi,\sigma}\approx 2^{-4qN}\delta_{\pi,\sigma} + O(2^{-4q(N-1)})$~\cite{Collins2006}. 
Thus, we retain only the dominant contributions in the sum~\eqref{eq:A8_twirled} with $\pi=\sigma$ and neglect the sub-leading terms. 
This approximation is already well motivated by the fidelity case: for two replicas, the off-diagonal Weingarten coefficients are suppressed by an additional factor of $2^{-N}$ relative to the diagonal ones (cf. Eq.~\eqref{eq:resN}), and an analogous suppression persists at higher replica number~\footnote{Normalized out-of-time-order correlators, which lie outside the scope of this work, instead require subleading contributions that must be treated using free-probability methods~\cite{pappalardi2022eigenstatethermalization,fava2025designs}.}.

The traces $\Tr\!\left(T_{\pi}\,(V_\alpha\otimes V_\alpha^\dagger)^{\otimes 4}\right)$ involved in Eq.~\eqref{eq:A8_twirled} can be evaluated analytically since both $T_\pi$ and $V_\alpha$ [see Eq.~\eqref{eq:coherent_error}] admit a product structure over $N$ sites of the system. Similarly, the contraction of the permutation operators $T_\sigma$ with the boundary operators in Eq.~\eqref{eq:B8_def} can be performed analytically since the latter also have a product structure over the $N$ qubits, see \textbf{App.}~\ref{app:onsite}. This allows us to find close expressions for the annealed SRE utilizing, $8$-copies and $12$-copies respectively in calculation of $\tilde{M}_2(\mathbf{0})$ and $\tilde{M}_3(\mathbf{0})$. Their full expressions at finite $N$ are reported in Ref.~\cite{OURzenodo}.

We start analyzing the $q=2$ case, for which a compact leading-order expression for the annealed SRE reads
\begin{equation}
    \tilde M_2 (\mathbf{0})= -\log_2 \left[ x^4 + 2^{2-k} (1-x)^2 (1 + 2x + 3x^2) \right]
    \label{eq:SRE2good}
\end{equation}
where $\gamma \equiv 2^{N-k} \cos^{2N}\left(\frac{\alpha}{2}\right)$ and $ x = \gamma/(1+\gamma)$. Eq.~\eqref{eq:SRE2good} is compared  with the quenched average $\overline{M}_2(\mathbf{0})$ in Fig.~\ref{fig:SRE2H}(a,c), highlighting close agreement of the numerically obtained quenched averages $\overline{M}_2(\mathbf{0})$ and the analytical expression for the annealed SRE $\tilde M_2 (\mathbf{0})$, which indirectly verifies the self-averaging of the numerator and denominator in \eqref{eq:sreAN1}. In passing, we note that the dominating contributions to $\tilde M_2 (\mathbf{0})$ are provided by permutation operators corresponding to \textit{noncrossing partitions}~\cite{fava2025designs} (this set has cardinality $\mathrm{C}_{4q}$ for $4q$ replicas with $\mathrm{C}_m$ the $m$-th Catalan number), and, additionally by genus one permutations~\cite{zuber2023counting, dowling2025freeindependenceunitarydesign}.

Close to $\alpha_c$, the SRE admits a particularly form as a function of system size $N$ and the error strength $\alpha$
\begin{equation}
\tilde M_2 (\mathbf{0})
=
4 \log_2 \left[ 1+ \exp\left(\tan(\alpha_c/2)\, N(\alpha-\alpha_c)\right) \right],
\label{eq:M2_haar_scaling}
\end{equation}
depending only on the single scaling variable $N(\alpha-\alpha_c)$. In the thermodynamic limit, this expression sharpens into a non-analytic transition at $\alpha=\alpha_c$: one has $\tilde M_2(\mathbf{0})\to 0$ for $\alpha<\alpha_c$, while for $\alpha>\alpha_c$ the SRE grows linearly as
$\tilde M_2(\mathbf{0})\simeq 4\tan(\alpha_c/2)\,N(\alpha-\alpha_c)$. Exactly at criticality, $\tilde M_2(\mathbf{0})=4$.
Therefore, the magic transition coincides with the encoding--decoding transition, while \eqref{eq:M2_haar_scaling} implies the that the correlation-length exponent is $\nu=1$.

We now specialize to $M_3$, and perform the $12$-replicas computation. Summing over $12!$ terms in the diagonal approximation of the sum~\eqref{eq:A8_twirled} (for $q=3$), we obtain the closed expression for the leading terms in $N\gg1$ limit
\begin{equation}
\tilde M_3 (\mathbf{0})  =-\frac{1}{2} \log \left[ x^6+2^{-k}{(1-x^2)(1+x^2+16x^4)}  \right]\;,
\label{eq:SRE3good}
\end{equation}
plotted in Fig.~\ref{fig:SRE2H}(b,d) against the numerical data for the quenched averaged of the SRE, $\overline M_3(\mathbf{0})$.
Our results strongly corroborate the analytical computation and highlight the similarity between $M_3$ and $M_2$. 
In particular, Eq.~\eqref{eq:SRE3good} has fully analgoous scaling form as Eq.~\eqref{eq:M2_haar_scaling}. The main difference is that the crossing point occurs at $M_3 = 3$ instead than $M_2=4$, and the scaling form is  $M_3 \approx 3 \log_2 \left[ 1 + \exp\left( N \tan\left(\frac{\alpha_c}{2}\right) (\alpha - \alpha_c) \right) \right] $. 

These analytical expression, combined with our numerical investigation for $q\ge 4$ which we do not report for presentation purposes, give the expression for the SRE density (implied by \eqref{eq:SRE2good} and \eqref{eq:SRE3good} for $q=2,3$, and conjectured for any $q\geq4$):
\begin{equation}
    \lim_{N \to \infty} \frac{M_q}{N} = 
    \begin{cases} 
          0 & \text{if } \alpha < \alpha_c \\[1em]
        \displaystyle \frac{4q}{q-1} \log_2\! \left[ \frac{\cos(\alpha_c/2)}{\cos(\alpha/2)} \right] & \text{if } \,\,
        2^{-\frac{r}{4q}} < \frac{\cos\left(\frac{\alpha}{2}\right)}{\cos\left(\frac{\alpha_c}{2}\right)} \le  1 \\[1em]
        \displaystyle \frac{r}{q-1} & \text{if } \frac{\cos\left(\frac{\alpha}{2}\right)}{\cos\left(\frac{\alpha_c}{2}\right)} \le 2^{-\frac{r}{4q}},
    \end{cases}\;
    \label{eq:SREden}
\end{equation}
shown in Fig.\ref{fig:SRE2}(c) for $q=2,3$.
Similarly to what happens for entanglement and participation entropy, cf. Ref.~\cite{turkeshi2024error} and \textbf{App.}~\ref{app:parti}, the magic phase transition is fixed by the error-resilience transition, with a multifractal phase at $\alpha\ge \alpha_c$ that depends explicitly on the R\'{e}nyi index $q$.

\begin{figure}
    \centering
    \includegraphics[width=1\columnwidth]{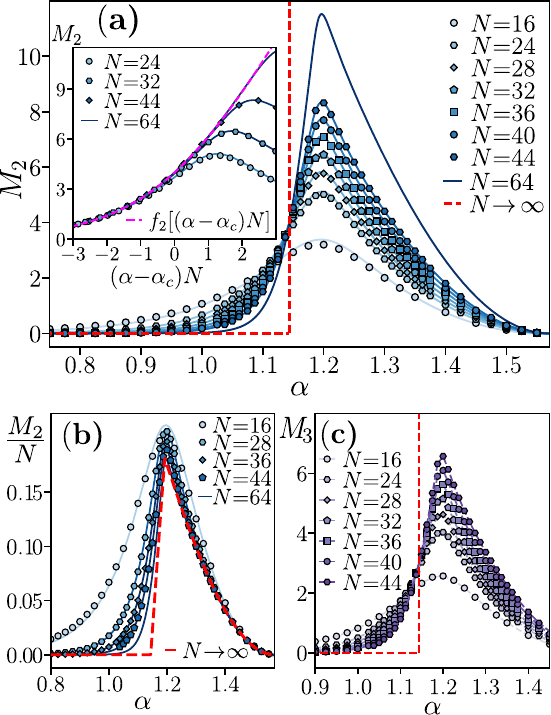}
    \caption{
Phase transition in magic resources for Clifford encoders with the postselected syndrome $\mathbf{s}=\mathbf{0}$. (a) The SRE $M_2$ as a function of the error strength $\alpha$ for various system sizes $N$ and a fixed code rate $r=k/N=1/2$. Solid lines represent the analytic formula~\eqref{eq:SRE2cliffgood}, while symbols denote the quenched averages $\overline{M}_2$ computed with Pauli propagation method [\textbf{Sec.}~\ref{sec:PauliPropaCliff}], averaged over more than 4000 circuit realizations. The inset demonstrates the collapse of $M_2$ onto the universal curve~\eqref{eq:M2_haar_scaling} (which coincides with the Haar case) when plotted against the rescaled error strength $(\alpha-\alpha_c)N^{1/\nu}$, with $\nu=1$. (b) The SRE density $M_2/N$, with red dashed lines indicating the $N\to\infty$ limit. (c) Numerically calculated quenched averages of the SRE $M_3$; dashed lines are shown to guide the eye.
    }
    \label{fig:CliffSRE}
\end{figure}

\subsubsection{Clifford encoders}
\label{subsec:maclif}
For Clifford encoders $\mathcal{E}=\mathcal{C}_N$, the calculation of the annealed SRE, Eq.~\eqref{eq:m2_annealed_replica} with Schur-Weyl duality requires many more terms than in the Haar case, with dimension that scales asymptotically as $O(2^{(n-2)^2/2})$ for $n$-replica systems. Nevertheless, for $q=2$, i.e. at $m=8$ replicas, the generalization of the expression in Eq.~\eqref{eq:Acli} still holds, with $\Omega, \Omega' \in \mathfrak{P}_8$.
Using the diagonal approximation of the $4q$-copy Clifford Weingarten matrix $\mathrm{Wg}^{\mathcal{C}_N}_{\Omega,\Omega'}\approx 2^{-4qN}\delta_{\Omega,\Omega'} + O(2^{-4qN-1)})$~\cite{Gross2021} which enables us to keep only the contributions from reduced Pauli monomials $\Omega = \Omega'$, we compute the leading contributions to the annelead average of the stabilizer entropy with R\'enyi index $q=2$ ($M_2$). 
The main simplification that enables calculating the sum over 9845550 elements of $\mathfrak{P}_8$ is to identify 13 classes of operators that split the $8$-copy Clifford commutant, cf. Ref.~\cite{Bittel25commutant}  and \textbf{App.}~\ref{App:averages} for details. The final ingredient enabling the fully analytic calculation is the representation $\Omega$ of the reduced Pauli monomial $\Omega$, which has a tensor product structure over the $N$ qubits.
Using symbolic calculus, we obtain the lengthy accurate expression for $\tilde{M}_2(\mathbf{0})$ for the Clifford encoders detailed in Ref.~\cite{OURzenodo}.

However, keeping the dominant terms in the expression enables us to find a compact closed form $\tilde{M}_2(\mathbf{0})$ that approximates well the exact results already for a systems comprising a few dozen of qubits.
Let us denote $ t \equiv \cos^2\!\left(\frac{\alpha}{2}\right)$, and introduce the following four nonnegative polynomials on $t\in[0,1]$:
\begin{equation}
  \begin{split}
B_1(t)&=16 t^4-32 t^3+20 t^2-4 t+1,\\
B_2(t)&=8 t^2-14 t+7,\\
B_3(t)&=4 t^2-6 t+3,\\
B_4(t)&=4 t^2-4 t+2.
\end{split}  
\end{equation}
Then, the compact expression for the annealed averagege of SRE for Clifford encoder is given by 
\begin{equation}
\tilde{M}_2 (\mathbf{0})
=
-\log_2\left[x^4
+x^2(1-x)^2\,\mathcal{Q}(t)
+(1-x)^4\,B_1(t)^N \right],
\label{eq:SRE2cliffgood}
\end{equation}
where we have defined the function
\begin{equation}
\mathcal{Q}(t)
=
2^{\,k+3-N}\,B_2(t)^N
+16\,B_3(t)^N
+7\,2^{\,1-k}\,B_4(t)^N.
\end{equation}
This enables the computation of the SRE density
\begin{equation}
    \lim_{N \to \infty} \! \frac{M_2}{N} \! = \!
    \begin{cases} 
        0 & \! \!\text{if } \alpha < \alpha_c \\[1em]
        \displaystyle 8 \log_2 \left[ \frac{\cos(\alpha_c/2)}{\cos(\alpha/2)} \right] & \! \! \! \! \! \! \! \! \!\text{if} \,\,\,
        \Lambda(\alpha) \!< \!\frac{\cos\left(\frac{\alpha}{2}\right)}{\cos\left(\frac{\alpha_c}{2}\right)} \!\le \!1
        \\[1em]
        \displaystyle -\log_2 \left[ 1 - \frac{1}{4}\sin^2(2\alpha) \right] & \! \!\text{if }\,\, \frac{\cos\left(\frac{\alpha}{2}\right)}{\cos\left(\frac{\alpha_c}{2}\right)} \le \Lambda(\alpha),
    \end{cases}\;\label{eq:jeez}
\end{equation}
where the boundary threshold $\Lambda(\alpha)$ is defined as:
\begin{equation}
    \Lambda(\alpha) = \left( 1 - \frac{1}{4}\sin^2(2\alpha) \right)^{\frac{1}{8}}
\end{equation}

The expressions in Eq.~\eqref{eq:SRE2cliffgood} and ~\eqref{eq:jeez} are
one of the central results of this work. 
As shown in Fig.~\ref{fig:CliffSRE}(a), Eq.~\eqref{eq:SRE2cliffgood} is robustly validated by the exact numerical data obtained for the quenched average of the SRE, $\overline{M}_2(\mathbf{0})$ for $N\le 44$.  Similarly, in Fig.~\ref{fig:CliffSRE}(b) we observe how the finite system results approach the asymptotic expression for the SRE density, Eq.~\eqref{eq:jeez}.
Importantly, the SRE once again presents the critical behavior same as for the Haar case, following the same scaling form, Eq.~\eqref{eq:M2_haar_scaling}. This demonstrates that the magic resource phase transition occurs at the same critical error strength $\alpha_c$ of the error-resilience phase transition and that the critical exponent is again $\nu=1$.

A closed formula for $q=3$ (requiring $12$-replicas) is prohibitive due to the grwoth of the dimension of the Clifford commutant, cf. Eq.~\eqref{eq:Sigma_k_cardinality}. Nevertheless, the numerical results obtained with the Pauli-propagation method, shown in Fig.~\ref{fig:CliffSRE}(c), demonstrate a qualitatively similar behavior for $M_3$ with respect to $M_2$, and in particular around the critical point $\alpha_c$. 
All in all these results demonstrate that, at fixed syndrome $\mathbf{s}=\mathbf{0}$ the transition in magic resources is fully captured by the error-resilience phase transition.

\subsection{Higher order syndromes}
\label{subsec:forced_clifford_magic}

We now discuss what happens when $\mathbf{s}\neq \mathbf{0}$ results are post-selected. 

\subsubsection{Haar encoders}
Up to this point, our analysis has focused on the error-free branch, characterized by the trivial syndrome measurement $\mathbf{s}=\mathbf{0}$. For Haar-random encoders, the concentration of measure ensures that all non-trivial syndrome measurements ($\mathbf{s}\neq \mathbf{0}$) effectively fall into a single, featureless equivalence class. Fixing a non-trivial syndrome result $\mathbf{s}\neq \mathbf{0}$, a direct computation for Eq.~\eqref{eq:pF_def} yields
\begin{equation}
    \mathbb{E}[p_U(\mathbf{s}\neq \mathbf{0})] = \frac{2^k(2^N-(1+\cos(\alpha))^N)}{2^{2N}-1} \simeq 2^{k-N}. 
\end{equation}
Evaluating the fluctuations around this mean gives $[\Delta p^\mathcal{U}(\mathbf{s}\neq \mathbf{0})]^2 \simeq 2^{-k}$, demonstrating that the denominator is asymptotically self-averaging. Following a similar procedure for the numerator, Eq.~\eqref{eq:mF_def}, we obtain
\begin{equation}
    \mathbb{E}[m_{F,U}(\mathbf{s}\neq \mathbf{0})] = \frac{2^N-(1+\cos(\alpha))^N}{2^{2N}-1} \simeq 2^{-N}. 
\end{equation}
Interestingly, the orthogonality condition $\langle \mathbf{s}|\mathbf{0}\rangle=0$ significantly simplifies the replica structure. The only non-vanishing contributions arise from permutations of the form $\sigma_{2q}\cdot \tau$, where $\tau\in \mathrm{S}_q\times \mathrm{S}_q$ is a factorized permutation. Employing the diagonal approximation, we find that the moments scale as:
\begin{equation}
    \mathbb{E}[(m_{F,U}(\mathbf{s}\neq \mathbf{0}))^q] \simeq q!\, \mathbb{E}[m_{F,U}(\mathbf{s}\neq \mathbf{0})] \simeq \frac{q!}{2^N}.
    \label{eq:facttt}
\end{equation}

The factorial growth of these moments admits a transparent probabilistic interpretation: it is characteristic of an exponential (Porter--Thomas) distribution, modulated by an overall activation probability $2^{-N}$. Equivalently, the random variable $\mu_F \equiv m_{F,U}(\mathbf{s}\neq \mathbf{0})$ is identically zero with probability $1-2^{-N}$ and exponentially distributed when conditioned on a nonzero outcome. Consequently, the full probability distribution of $\mu_F$ is given by
\begin{equation}
    P(\mu) = (1-2^{-N})\,\delta(\mu) + 2^{-N} e^{-\mu}\,\Theta(\mu),
\end{equation}
which precisely reproduces the moment structure of Eq.~\eqref{eq:facttt} in the scaling limit $N\to\infty$ for $0<\alpha<\pi/2$. Assembling these results, we conclude that the fidelity converges uniformly to the constant random-state value of $2^{-k}$ for any $\alpha$. Thus, in the $\mathbf{s} \neq \mathbf{0}$ branches, the system is featureless and the error-resilience phase transition is provably absent for Haar encoders.

\begin{figure*}
    \centering
    \includegraphics[width=0.75\linewidth]{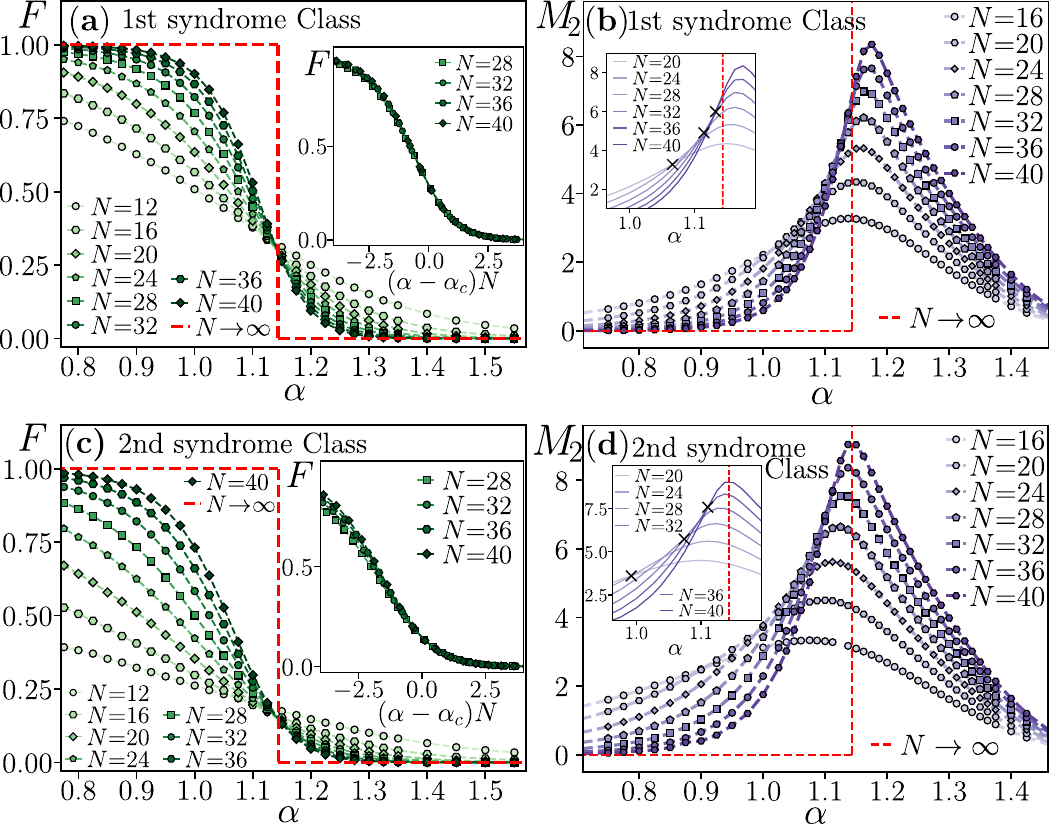}
    \caption{
Phase transition in magic resources for Clifford encoders at a fixed code rate $r=k/N=1/2$ across varying syndrome classes. The fidelity $F$ as a function of the error strength $\alpha$ is shown for the first and second syndrome classes in panels (a) and (c), respectively. The SRE $M_2$ as a function of $\alpha$ undergoes a transition between area-law and volume-law phases at $\alpha=\alpha_c$, shown for the first and second syndrome classes in panels (b) and (d), respectively. The insets in (a) and (c) demonstrate the collapse of the fidelity $F$ upon rescaling $\alpha \mapsto (\alpha-\alpha_c)N^{1/\nu}$ (with $\nu=1$). Meanwhile, the insets in (b) and (d) highlight the crossing point in the $M_2$ versus $\alpha$ curves, which converges to $\alpha=\alpha_c$ with increasing system size $N$. Results are computed  with Pauli propagation [\textbf{Sec.}~\ref{sec:PauliPropaCliff}] and averaged over more than 2000 circuit realizations.
    }
    \label{fig:SRE2}
\end{figure*}

\subsubsection{Higher syndrome classes for Clifford encoders}
\label{subsec:higher}

This situation is drastically different for the Clifford case, which retains significantly more structure. By direct inspection, one finds that the relative fluctuations of the Born probabilities grow exponentially with system size, $[\Delta p^\mathcal{C}(\mathbf{s}\neq \mathbf{0})]^2 \simeq \mathcal{O}(\exp(N))$. This exponential growth indicates a strong breakdown of self-averaging. Because the analytical computation leads to strictly non-self-averaging behavior when $\mathbf{s}\neq \mathbf{0}$, the annealed approximation is rendered invalid. 
Here, instead, to understand the physics of the encoding-decoding circuits with Clifford encoders for $\mathbf{s}\neq \mathbf{0}$, we utilize the insight from Eq.~\eqref{eq:PauliProp} about the emergence of syndrome classes for Clifford encoders. 

In the following investigation, we fix a syndrome class, i.e., for given Clifford unitary $C=U$, we consider the logical state $\ket{\psi_U(\mathbf{s})}_L$ only for syndromes $\mathbf{s}$ of a fixed class $\ell_{\mathbf{s}}$. We note that the class of a selected syndrome $\mathbf{s}$ changes from one unitary $C$ to another. Hence, for each fixed encoder $C$, we specifically select the syndromes of a fixed class $\ell_{\mathbf{s}}=1,2$ and evaluate the quenched averages of the quantities of interest for the corresponding logical states $\ket{\psi_U(\mathbf{s})}_L$. 

The discussion in \textbf{Sec.}~\ref{sec:PauliPropaCliff}, for the initial state $\ket{\mathbf{0}}_L\otimes \ket{\mathbf{0}}_S$, implies that the dominating contribution to the final logical state for the syndrome $\mathbf{s}$ is 
\begin{equation}
    \ket{\psi_d}_L \propto \delta_{\mathbf{s},\,\mathbf{y}^A}\ket{\mathbf{x}^A},
\end{equation}
where $|A| =\ell_{\mathbf{s}}$ is the class of the syndrome $\mathbf{s}$. In the error-free branch, $\mathbf{s}=\mathbf{0}$, the dominating contribution is proportional to the initial logical state $\ket{\mathbf{0}}_L$. In contrast, for $\ell_{\mathbf{s}} \geq 1$, the dominating computational basis state $\ket{\mathbf{x}^A}$ is determined by the syndrome $\mathbf{s}$, and the Clifford unitary $C$ via Eqs.~\eqref{eq:pauliconj}-\eqref{eq:oversn}. 
Consequently, for $\ell_{\mathbf{s}}\geq1$, the fidelity, Eq.~\eqref{eq:Fs} is measured with respect to the state $\ket{\mathbf{x}^A}$:
\begin{equation}
F_U(\mathbf{s}) = |\langle \mathbf{x}^A | \psi_{U}(\mathbf{s}) \rangle|^2.
\label{eq:Foffdiag}
\end{equation}
Additionally, to quantify magic resources of the decoded logical state, we calculate the quenched averaged stabilizer entropy.

The results for the logical fidelity $F$ and the SRE $M_2$ are shown in Fig.~\ref{fig:SRE2}(a,b) for the syndromes of the first class $\ell_{\mathbf{s}}=1$, and in Fig.~\ref{fig:SRE2}(c,d) for the second class syndromes, $\ell_{\mathbf{s}}=2$.
The fidelity results for both $\ell_\mathbf{s}=1$ and $\ell_\mathbf{s}=2$ exhibit a clear signature of the error-resilience phase transition, with fidelity approaching unity for $\alpha < \alpha_c = \arccos( 2^{\frac{r-1}{2}})$, and $F\to 0$ for $\alpha > \alpha_c$. While the value of $F$ at the critical point $\alpha=\alpha_c$ decreases with increasing syndrome class $\ell_\mathbf{s}$, the results collapse on the universal function of $(\alpha- \alpha_c)N$, showing the universality class of the transition ($\nu=1$) for both syndrome classes is the same as in the $\mathbf{s}=\mathbf{0}$ (i.e. $\ell_\mathbf{s}=0$) case.

The SRE results for $\ell_\mathbf{s}=1$ and $\ell_\mathbf{s}=2$ syndrome classes show that the error-resilience phase transition observed on the level of logical state fidelity~\eqref{eq:Foffdiag}, is also manifested in the magic state resources behavior. For $\alpha < \alpha_c$, we observe a monotonic decrease of $M_2$. The crossing points, shown in the insets in Fig.~\ref{fig:SRE2}(b,d), approach the critical point $\alpha_c$ with increasing system size $N$. While the accessible system sizes are insufficient to demonstrate scaling collapses of the SRE upon rescaling $\alpha \mapsto (\alpha-\alpha_c)N$, the results for $\ell_{\mathbf{s}}=1,2$ are consistent with a transition similar to the $\ell_{\mathbf{s}}=0$ case, between a phase at $\alpha < \alpha_c$ with vanishing SRE density and a phase with a finite SRE density for $\alpha > \alpha_c$.

These results show that the magic resource phase transition in encoding-decoding circuits with Clifford encoders is a phenomenon extending over the error-free branch $\mathbf{s}=0$. Importantly, in the error protecting phase for $\alpha < \alpha_c$, the logical fidelity [cf. Eq.~\eqref{eq:Foffdiag}]  increases to unity with increasing system size $N$, showing a perfect recovery of the state $\ket{\mathbf{x}^A}$, obtained by a single Pauli string $P_A$ action on the initial state. The latter reduces only to local qubit rotations, does not produce entanglement, and can be undone, or \textit{corrected}, by the action of $P_A$ itself. In this sense, the error-protecting phase, and the related phase with vanishing SRE, generalize to arbitrary syndrome classes $\ell_{\mathbf{s}}$ for encoding-decoding circuits with Clifford encoders.
In \textbf{App.}~\ref{app:heuristic_magic_bound}, we further corroborate this conclusion by deriving an explicit bound on the stabilizer entropy in terms of the fidelity for an initial logical stabilizer state.

\section{Born-Rule Measurements}
\label{sec:born}
In experimental realizations of monitored quantum circuits, post-selecting on a specific syndrome outcome $\mathbf{s}=\mathbf{0}$ incurs an exponential in the number $N-k$ of ancillary qubits overhead, rendering the ``forced'' measurement ensemble inaccessible for experiments at large system sizes.
Instead, experiments naturally sample the ``Born-rule'' ensemble, where the syndrome $\mathbf{s}$ is observed with probability 
$p_{U}(\mathbf{s})=\Tr[\Pi_{I,\mathbf{s}}\rho_U]$.
For notational simplicity, we consider the logical pure state $\rho_{L,U}(\mathbf{s})\equiv\ket{\psi_{U}(\mathbf{s})}\!\bra{\psi_{U}(\mathbf{s})}$. 
Then, the physical quantity of interest $\mathcal{O}(|\psi\rangle)$, such as the fidelity or the stabilizer entropy, therefore, becomes the weighted average over all possible measurement trajectories
\begin{equation}
    \overline{\mathcal{O}}_{\text{BR}} = \mathbb{E}_{U\sim \mathcal{E}} \left[ \sum_{\mathbf{s} \in \mathbb{Z}_2^{N-k}} p_U(\mathbf{s}) \mathcal{O} (| \psi_{U}(\mathbf{s}) \rangle) \right].
    \label{eq:Born}
\end{equation}
For example, the fidelity~\eqref{eq:Foffdiag} with respect to the dominating computational basis state $\ket{\mathbf{x}^A}$
is the Born-rule sampled average 
\begin{equation}
    \overline{F}_{\mathrm{BR}} \equiv \mathbb{E}_{U\sim \mathcal{E}}\sum_{\mathbf{s} \in \mathbb{Z}_2^{N-k}} p_{U}(\mathbf{s})\,F_U(\mathbf{s})\;. 
    \label{eq:Favg}
\end{equation}
Analogously, we define also the Born-rule sampled average of the SRE, $M^{\mathrm{BR}}_2 = \mathbb{E}_{U\sim \mathcal{E}}\sum_{\mathbf{s} } p_{U}(\mathbf{s}) M_2 (\rho_{L,U}(\mathbf{s}))$.  
As we shall see in the following, the Born probability $p_{U}(\mathbf{s})$ introduces a competition with the properties of the conditional logical state $\rho_{L,U}(\mathbf{s})$ and the syndrome class $\ell_{\mathbf{s}}$, leading to a critical phenomena different than in the post-selected cases.

\subsection{Haar Encoders: Instability of the Transition}
\label{subsec:born_haar}

For Haar-random encoders, the distinction between the forced (post-selected) and Born-rule ensembles is suppressed by the concentration of measure. 
As previously discussed, in the thermodynamic limit $N \to \infty$, the probability distribution of syndromes $p_U(\mathbf{s})$ becomes sharply concentrated around the uniform value $p_U(\mathbf{s}) \approx 2^{-(N-k)}$. 
For any syndrome $\mathbf{s}\neq \mathbf{0}$ we argued the error-resilience transition is no longer present. Thus, the weighted averages of the fidelity and stabilizer entropy exhibit no sharp transitions, only a smooth dependence on the error strength $\alpha$.
Thus, for Haar encoders, the only non-trivial transition occurs, only in the forced measurement case, when $\mathbf{s}=\mathbf{0}$.

\begin{figure}
    \centering
    \includegraphics[width=0.9\columnwidth]{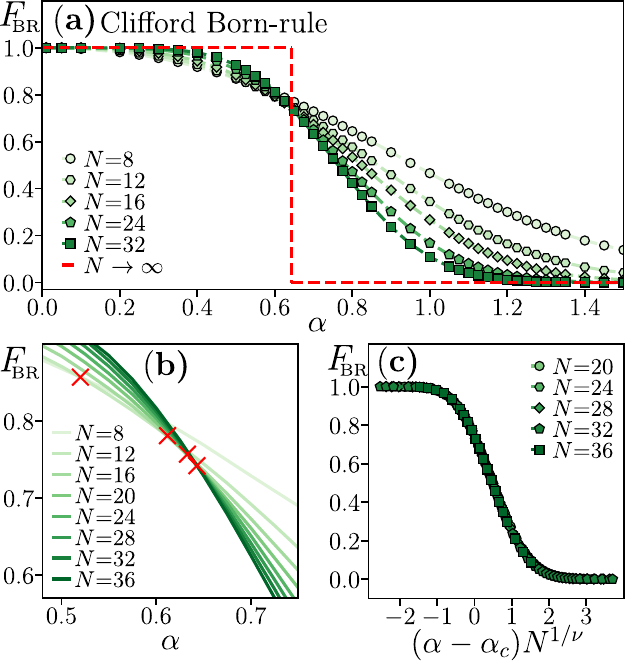}
    \caption{
Error-resilience phase transition in Clifford encoders with Born-rule syndrome averaging at a fixed code rate $r = k/N = 1/2$. (a) The fidelity $F_{\mathrm{BR}}$ as a function of the error strength $\alpha$ approaches a step-like behavior with increasing system size $N$. (b) The crossing point $\alpha_x$ of the $F_{\mathrm{BR}}$ curves shifts at small $N$, moving from $\alpha_x=0.52(1)$ (for $N=8, 12$) to $\alpha_x=0.633(3)$ (for $N=16, 20$), and stabilizes at $\alpha^{\mathrm{BR}}_c= 0.643(3)$ for $N\geq 20$. (c) The fidelity collapses onto a system-size-independent curve upon rescaling $\alpha\mapsto (\alpha-\alpha^{\mathrm{BR}}_c )N^{1/\nu}$, with the critical exponent $\nu=2$. Results are computed using the Pauli propagation technique and averaged over more than 1000 circuit realizations.
}
    \label{fig:FideAve}
\end{figure}

\subsection{Clifford Encoders: Shifted Criticality}
\label{subsec:born_clifford}

The scenario is fundamentally different for Clifford encoders. The Clifford group forms only a unitary 3-design for qubits, which is insufficient to concentrate the higher moments of the probability distribution $p_U(\mathbf{s})$. As a result, this probability fluctuates significantly depending on the choice of the syndrome $\mathbf{s}$ for a specific Clifford unitary $C$~\cite{magni2025anticoncentration}. Consequently, different syndromes project the system into distinct classes of states with vastly different properties and magic content, as observed already in \textbf{Sec.}~\ref{subsec:higher}. This lack of concentration leads to a markedly richer phenomenology for non-trivial syndromes ($\mathbf{s}\neq \mathbf{0}$) than in the Haar case. To probe these phenomenology, we use the Pauli propagation method to evaluate the Born-rule averaged $\overline{F}_{\mathrm{BR}}$ and $M^{\mathrm{BR}}_2$.

The most striking consequence of Born-rule sampling is the displacement of the phase transition. Let us first consider the logical fidelity $\overline{F}_{\mathrm{BR}}$, as shown in Fig.~\ref{fig:FideAve}. The critical error strength for the transition shifts to a lower value, $\alpha_c^{\text{BR}} \approx 0.65$, compared to the intrinsic error-resilience threshold $\alpha_c$ observed in the forced case, cf. \textbf{Sec.}~\ref{subsubsec:forced_cliff} and \ref{subsec:higher}. This shift arises because 
the non-trivial syndromes $\mathbf{s} \neq \mathbf{0}$---those associated with a degraded logical state---acquire significant probability weight $p_U(\mathbf{s})$ even before the code's theoretical capacity is exceeded. The syndrome probability itself develops a scaling behavior that is inequivalent to, and competes with, the error-resilience scaling in each fixed syndrome class $\ell_{\mathbf{s}}$.

Furthermore, the transition in the Born ensemble is sharpens more slowly than its forced-measurement counterpart. The extensive statistical fluctuations of $p_U(\mathbf{s})$ act as a relevant perturbation to the critical point, leading to stronger finite-size drifts and fundamentally modifying the critical scaling. Crucially, the correlation length critical exponent extracted through finite-size scaling collapses from our numerical results is compatible with $\nu=2$, see Fig.~\ref{fig:FideAve}(c). This highlights a clear inequivalence of universality classes between the forced and Born-rule transitions in encoding-decoding circuits with Clifford encoders.

This mechanism naturally explains the discrepancy between the clean, $\nu=1$, universality class derived from pure error-correction principles (\textbf{Sec.}~\ref{sec:forced}) and the non-universal exponents reported in prior numerical studies of monitored Clifford circuits~\cite{Niroula2024phasetransition}. The experimentally observed transition is a dressed phenomenon: it is the fundamental error-resilience transition obscured and modified by the sampling statistics of the decoder.

\begin{figure}
    \centering
    \includegraphics[width=1\columnwidth]{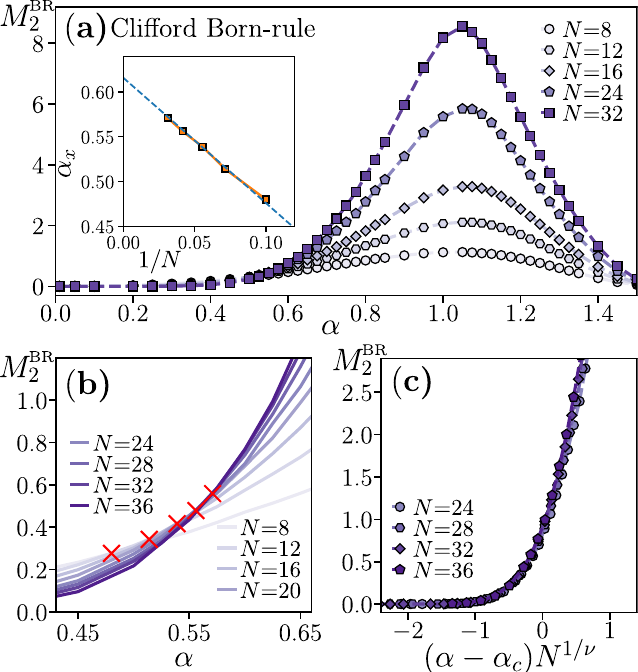}
    \caption{
Phase transition in magic resources for Clifford encoders with Born-rule syndrome averaging at a fixed code rate $r=k/N=1/2$. (a) The SRE $M^{\mathrm{BR}}_2$ as a function of the error strength $\alpha$. The inset shows the crossing point (displayed in panel (b)) as a function of $1/N$, alongside an extrapolation to the large-$N$ limit, where $\alpha_x \to 0.616 = \alpha_s \approx \alpha_c^{\mathrm{BR}}$. (c) The collapse of the SRE when plotted as a function of the rescaled error strength $(\alpha-\alpha_s)N^{1/\nu}$, with the critical exponent $\nu=2$. Results are computed using the Pauli propagation technique and averaged over more than 1000 circuit realizations.
}
    \label{fig:SREAve}
\end{figure}

A fully analogous phenomenology holds for the magic resources quantified by the stabilizer entropy $M^{\mathrm{BR}
}_2$, as shown in Fig.~\ref{fig:SREAve}. Once again, the Born-rule probability induces a modified scaling behavior, with the transition shifting to $\alpha_c^{\text{BR}} \simeq 0.62$ and a critical exponent of $\nu=2$. These results are entirely compatible with those of the fidelity transition demonstrating that also in the Born-rule case, the phase in which the initial logical state is not recovered is associated with appearance of extensive magic resources. Altogether, the results for $\overline{F}_{\mathrm{BR}}$,  $M^{\mathrm{BR}
}_2$, supported by the participation entropy (see \textbf{Appendix}~\ref{app:parti}), underscore that a fundamentally different mechanism---driven by the syndrome sampling statistics---governs the criticality of the system in the Born-rule ensemble.

\subsection{Discussion}
\label{sec:discussion}

The results presented in this work provide a unified explanation of magic phase transitions in encoding--decoding quantum circuits.
We have shown that, for forced measurements, Clifford encoders exhibit a magic phase transition that is identical for \emph{any fixed syndrome error class} $\ell_{\mathbf{s}}$.
By contrast, when Born-rule sampling is considered, the physics is fundamentally modified by the inclusion of the syndrome probabilities  $p_{U}(\mathbf{s})$.

A central observation is the crucial role of the replica limit in capturing the correct physics both for forced measurements in nontrivial syndrome classes $\ell_{\mathbf{s}\neq \mathbf{0}}$ and for Born-rule–weighted observables.
In these situations, self-averaging fails: fluctuations of the relevant quantities remain finite in the thermodynamic limit, and annealed averages no longer provide a faithful description.
As a consequence, a replica trick is required, introducing auxiliary replicas to correctly account for probability weights and logarithmic observables.

For instance, the Born-rule–averaged second stabilizer R\'enyi entropy can be written schematically as
\begin{equation}
\begin{split}
M_2^{\mathrm{BR}}
=
-\lim_{n\to 0}\frac{1}{n}\,
\mathbb{E}_{U\in\mathcal{E}}
\left[
\sum_{\mathbf{s}} p_U(\mathbf{s})\,
\bigl( m_{M,U}^{(2)}(\mathbf{s})^{\,n} - 1 \bigr)
\right]
\\
+\lim_{n\to 0}\frac{1}{n}\,
\mathbb{E}_{U\in\mathcal{E}}
\left[
\sum_{\mathbf{s}} p_U(\mathbf{s})\,
\bigl( p_U(\mathbf{s})^{\,2n} - 1 \bigr)
\right],
\label{eq:replicaSRE}
\end{split}
\end{equation}
where the logarithms entering the definition of the stabilizer R\'{e}nyi entropy are generated by taking the replica limit $n\to 0$.
At finite integer $n$, all quantities appearing in the averages are polynomial functions of the encoding unitary $U$, allowing their evaluation via standard replica methods.
A full calculation of $M_2^{\mathrm{BR}}$ would then require performing an analytic continutation of the finite $n$ result to the $n\to 0$ limit.

When measurements are \emph{forced} to the trivial syndrome $\mathbf{s}=0$, concentration of measure ensures self-averaging: fluctuations are exponentially suppressed with system size, and the replica limit $n \to 0$ becomes unnecessary
In this case, the annealed averages correctly capture the physical behavior of encoding-decoding circuits.

In contrast, the other situations considered in this work --- namely forced measurements in $\ell_{\mathbf{s}}\geq 1$ syndrome classes and the Born-rule sampling --- go beyond this framework. The problem of deciding a class $\ell_{\mathbf{s}}$ of a syndrome $\mathbf{s}$ for a given Clifford encoder $C$ is non-linear, therefore evading the replica treatment. Nevertheless, our Pauli propagation scheme results indicate that the transition in fixed syndrome classes $\ell_{\mathbf{s}}$ belong to the same universality class as in the error-free $\mathbf{s}=\mathbf{0}$ branch.
In the Born-rule sampling case, the fluctuations between the syndrome probabilities are provably large and the replica limit, Eq.~\eqref{eq:replicaSRE}, is essential to capture the magic resources phase transition.
This limit significantly changes the position and the critical properties of the transition, underlying the different universality classes identified in this work.

\section{Conclusion}
\label{sec:conclusion}

\subsection{Summary}
In this work, we have developed a unified theory of magic phase transitions in encoding--decoding quantum circuits, combining exact analytical results with extensive numerical simulations.
In the forced-measurement regime, we demonstrated that the magic phase transition is directly tied to the underlying error-resilience transition of the code~\cite{turkeshi2024error}. For Haar-random encoders, this transition occurs exclusively in the success branch of the decoder, $\mathbf{s}=\mathbf{0}$, where we obtain the exact critical point and correlation-length exponent $\nu=1$.
For the more structured case of Clifford encoders, we showed that an analogous transition arises within any fixed syndrome class $\ell_{\mathbf{s}}$. In this regime, the phenomenology of the magic transition is identical to that of the success branch, with the same critical point and universality class as in the Haar case.

By contrast, when syndrome outcomes are sampled according to the Born rule, we identified a distinct critical phenomenon driven by the probability distribution of the syndrome outcomes, irrespective of their class.
This effect acts as a relevant perturbation in the renormalization-group sense and induces a finite-size crossover toward a new critical point $\alpha_c^{\mathrm{BR}}$, which is shifted to lower error strength compared to the error-resilience threshold and is characterized by a correlation-length exponent $\nu=2$.

Taken together, these results resolve the ambiguity in the critical scaling and the multiple critical exponents observed for magic phase transitions in Clifford encoding--decoding circuits~\cite{Niroula2024phasetransition}, and highlight the central role of the replica limit in the statistical-mechanics description of monitored quantum systems with Born-rule measurements.
Furthermore, our results exemplify a transition in magic-state resources driven by the interplay of unitary dynamics and measurements, aligning with recent observations in hybrid quantum circuits~\cite{bejan2024dynamical,fux2024entanglement,paviglianiti2024estimatingnonstabilizernessdynamicssimulating}. In encoding-decoding circuits, our results demonstrate analytically that this transition in magic is fundamentally linked to the recoverability of the initial state, thereby coinciding with the entanglement transition between area- and volume-law phases~\cite{turkeshi2024error}.

\subsection{Outlook}
Our framework establishes a direct connection between error resilience and the dynamics of non-stabilizerness, opening several concrete directions for future research.

\paragraph{Interplay with incoherent noise.}
While our analysis focused on coherent errors, an important open problem is to understand how magic transitions generalize to genuinely open quantum systems.
Recent work~\cite{Trigueros25nonstab} has examined encoding--decoding circuits subject to non-unital incoherent noise, such as amplitude damping, and reported a striking qualitative difference with respect to the coherent case.
In particular, they found that the transition in decoding fidelity can decouple from the magic transition: dissipative processes may generate non-stabilizerness even when logical information is no longer recoverable.
This observation indicates that incoherent noise acts through a fundamentally different mechanism, breaking the tight correspondence between error resilience and magic generation established here for coherent errors.
Elucidating the microscopic origin of this decoupling, and determining whether it reflects a distinct universality class or a crossover phenomenon, remains an important open direction.

\paragraph{Relationship between entanglement and magic resources.}
The interplay between entanglement and magic has been widely explored, motivating numerical and analytical approaches that combine entanglement-based techniques, such as tensor networks, with resource-theoretic diagnostics of non-stabilizerness~\cite{tirrito2024quantifying,viscardi2025interplayentanglementstructuresstabilizer,hang2025,qian2025quantum,szombathy2025independentstabilizerrenyientropy,korbany2025longrangenonstabilizernessphasesmatter,tarabunga2025efficientmutualmagicmagic,cao2025gravitationalbackreactionmagical,qiant2024a,qiant25,qian2025,huang2024cliffordcircuitsaugmentedmatrix,yosprakob2026cliffordcircuitsaugmentedgrassmann}.
In these approaches, Clifford or classically simulatable circuits are used to \textit{disentangle} the many-body state, while the remaining magic is compressed into a low-entanglement representation, typically a matrix product state~\cite{masotllima2024stabilizer,masotllima2026limitsclifforddisentanglingtensor,fan2025disentangling,frau2025,fux2025disentanglingunitarydynamicsclassically}.

Our results naturally connect to recent efforts in this direction. 
In the encoding--decoding setting studied here, syndrome measurements disentangle the logical subspace from the environment while suppressing, or, beyond threshold, reshaping, the magic injected by coherent errors.
This demonstrates that entanglement removal and magic purification are distinct processes that can, in principle, be independently controlled~\cite{aziz2025classicalsimulationslowmagic}.
Exploiting this separation through active Clifford feedback and adaptive decoding may enable new strategies for magic-state distillation.

\paragraph{Non-Ergodic Dynamics and Many-Body Relevance.}
Framing the magic transition as a manifestation of error resilience provides a new lens to study the dynamics of non-stabilizerness in generic Hamiltonian systems~\cite{zhang2025stabilizerrenyientropytransition,jasser2025stabilizerentropyentanglementcomplexity,tirrito2025anticoncentrationnonstabilizernessspreadingergodic,moca2025nonstabilizernessgenerationmultiparticlequantum,bera2025nonstabilizernesssachdevyekitaevmodel,falcão2025magicdynamicsmanybodylocalized,sticlet2025nonstabilizernessopenxxzspin,tirrito2025universalspreadingnonstabilizernessquantum,hoshino2025stabilizerrenyientropyconformal,Haug23monotones,sun2026connectingmagicdynamicsthermofield}.
In non-ergodic regimes, such as many-body localized phases~\cite{Aba19col, Sierant25mbl} or systems with fragmented Hilbert spaces~\cite{Sala2020,Khemani2020}, local integrals of motion naturally protect subsets of logical degrees of freedom from global scrambling.
Our framework suggests that such systems may host intrinsically robust, error-resilient phases of magic, providing a genuinely non-local and resource-theoretic diagnostic of ergodicity breaking in many-body physics. 
Finally, extending the encoding--decoding paradigm to other quantum resources, such as non-Gaussianity~\cite{sierant2025,Bittel25ff, Falcao2026faf, paviglianiti2026emergence}, represents an intriguing direction for future work.

\paragraph{Relation with the Hayden--Preskill protocol.}
Our encoding--decoding architecture is closely related to the Hayden--Preskill black hole information recovery protocol~\cite{HaydenPreskill, YoshidaKitaev, RamppClaeys, Leone, lo1,lo2}. 
In the Hayden--Preskill setting, a Haar-random unitary models the scrambling dynamics of a black hole, while the collection of Hawking radiation plays the role of an information extraction channel. Successful decoding is possible only once the collected radiation exceeds a critical fraction, marking a transition to effective quantum error-correcting behavior.

Within this correspondence, the unitary encoder in our circuit implements the scrambling dynamics, whereas the syndrome measurement and decoding emulate the recovery operation performed by an external observer. Our results show that this decoding transition has a direct counterpart at the level of magic resources: the ability to recover the logical information is inseparable from the ability to recover non-stabilizer structure in the decoded state. 
In particular, below the error-resilience threshold, the decoding fails and the logical state is projected onto the stabilizer manifold, implying that any magic injected during the evolution is irreversibly lost to the environment. Only once the system enters the error-vulnerable phase, equivalently, beyond the Hayden--Preskill decoding threshold, does extensive magic survive in the logical subspace. In this sense, the computational power encoded in magic becomes accessible to the outside observer if and only if the information recovery protocol succeeds.

\paragraph{When errors make magic resources.}
At a broader level, our results suggest a shift in perspective on the role of measurements and noise in many-body quantum systems~\cite{Turkeshi2024coherent,Zihan2025}.
Rather than viewing measurements solely as a mechanism to suppress complexity, the error-vulnerable phase reveals that measurements can be used \emph{actively} to generate quantum resources.
In particular, once error correction fails, coherent errors combined with syndrome measurements inject extensive magic directly into the logical state.
The magic transition therefore marks not only the loss of error resilience, but also the onset of a regime in which measurements act as a resource-generating process.

The central challenge left open by this work is how to control and exploit the magic resource thus generated.
Understanding whether such resources can be structured, steered, or distilled, rather than appearing in an uncontrolled form, remains an important direction for future research.

\section*{Acknowledgements}
We thank Lorenzo Leone for discussions about the Clifford commutant and C. White for comments.
We thank B. Magni, D. Iannotti and F. Ballar Trigueros for comments on the manuscript.
X.T. acknowledges support from DFG under Germany's Excellence Strategy – Cluster of Excellence Matter and Light for Quantum Computing (ML4Q) EXC 2004/1 – 390534769, and DFG Collaborative Research Center (CRC) 183 Project No. 277101999 - project B01.
P.S. acknowledges fellowship within the “Generación D” initiative, Red.es, Ministerio para la Transformación Digital y de la Función Pública, for talent attraction (C005/24-ED CV1), funded by the European Union NextGenerationEU funds, through PRTR.

\section*{Code and Data Availability}
Code and Data of this project will be shared publicly at publication.

\appendix

\section{Participation entropy}
\label{app:parti}
We characterize the spreading of the wavefunction in the computational basis using the \textit{participation entropy} (PE)~\cite{Liu25,sierant2022universalbehaviorbeyond,Turkeshi_2024_hiler,lami2025anticoncentration,mace2019multifractal,luitz2014universalbehaviorbeyond}
\begin{equation}
    \mathrm{S}_q( \rho_{L,U}) = \frac{1}{1-q} \log_2 \left( \sum_{\textbf{x} \in \mathbb{Z}_2^k} \langle \textbf{x} | \rho_{L,U}(\textbf{s}) | \textbf{x} \rangle^q \right).
\end{equation}
Similarly to the SRE, the form of the logical state $\rho_{L,U}(\mathbf{s}$, ~\eqref{eq:rhoL}, leads to an expression involving a ratio
\begin{equation}
    \mathrm{S}_q( \rho_{L,U}) = \frac{1}{1-q} \log_2 \left( \sum_{x \in \mathbb{Z}_2^k} \frac{m^{(q)}_{S,U} (\mathbf{s})}{p_U^q(\mathbf{s})}\right),
\end{equation}
where the denominator $p_U(\mathbf{s})$ is given by Eq.~\eqref{eq:pF_def}, and the numerator is
\begin{equation}
    m^{(q)}_{S,U}(\mathbf{s}) = \sum_{x \in \mathbb{Z}_2^k } \bra{\Psi_0}\,U^\dagger V_\alpha^\dagger U\,\Pi_{x,\mathbf{s}}\,U^\dagger V_\alpha U\,\ket{\Psi_0}^{q},
    \label{eq:srePE2}
\end{equation}
where $\Pi_{x,\mathbf{s}} = (|x \rangle\langle x| )_L \otimes (| \mathbf{s} \rangle\langle \mathbf{s} |)_S$.
Throughout this section, we focus on the forced measurements with $\mathbf{s}=\mathbf{0}$. 

The \textit{quench PE} is the average $\overline{S}_q = \mathbb{E}[{S}_q(\rho_{L,U})]$. 
In analogy with the stabilizer entropy, we introduce the \textit{annealed PE} as 
\begin{equation}
    \tilde{S}_q( \mathbf{0}) = \frac{1}{1-q} \log_2 \left( \sum_{x \in \mathbb{Z}_2^k} \frac{ \mathbb{E}_{U\sim\mathcal{E}}[m^{(q)}_{S,U}  (\mathbf{0})]}{\mathbb{E}_{U\sim\mathcal{E}}[p_U^q(\mathbf{0})]}\right),
    \label{eq:PEann}
\end{equation}
and observe that its calculation reduces to 
\begin{equation}
    \tilde{S}_q(\mathbf{0}) 
= - \log_2 \left( 
\frac{
\Tr\!\left(
\mathcal{B}^{(\mathrm{S}_q)}_{\mathrm{num}}\;
\mathbb{E}_{U\sim\mathcal{E}}\!\left[A_U^{(2q)}\right]
\right) 
}{
\Tr\!\left(
\mathcal{B}^{(\mathrm{S}_q)}_{\mathrm{den}}\;
\mathbb{E}_{U\sim\mathcal{E}}\!\left[A_U^{(2q)}\right]
\right) 
}\right),
\label{eq:pe_annealed_replica}
\end{equation}
with $A_U^{(2q)}$ given in Eq.~\eqref{eq:A8_def}, and the replica boundary operators
\begin{equation}
\begin{split}
\mathcal{B}^{(\mathrm{S}_q)}_{\mathrm{num}}&\equiv T_{\sigma_{2q}} \!
\sum_{x \in \mathbb{Z}_2^k} \bigl[\mathcal{B}( \ket{x}\!\bra{x})\bigr]^{\otimes q},
\\
\mathcal{B}^{(\mathrm{S}_q)}_{\mathrm{den}}&\equiv T_{\sigma_{2q} } 
\bigl[\mathcal{B}(I)\bigr]^{\otimes q}\;,
\label{eq:Bpe_def}
\end{split}
\end{equation}
where we recall $\sigma_{2q}=(12)(34)\cdots(2q-1\ 2q)$.
As shown in \textbf{App.}~\ref{app:onsite}, these boundary operators can be conveniently written as a product over sites~\cite{Turkeshi_2023}, enabling symbolic calculations of annealed average of the participation entropy both for Clifford and Haar encoders.

For $q=2$ and $q=3$ we can use the Clifford Schur-Weyl duality to obtain the asymptotic closed expressions
\begin{align}
    \mathrm{S}_2&= -\log_2\left[ x^2 + \frac{(1-x)^2}{2^k} \left( B^N + 2 \right) \right] 
\label{eq:P2good}\\
S_3 &= -\frac{1}{2}\log_2 \left[x^3 + \frac{(1-x)^3}{2^{2k}} \left( A^N + 9B^N + 12 \right) \right]
\label{eq:S3good}
\end{align}
where we introduced the parameter $\gamma \equiv 2^{N-k} \cos^{2N}\left(\alpha/2\right)$ and the scaling constant $ x = \gamma/(1+\gamma)$, and defined for convenience $A  =1+3\cos^2\alpha$, $B =1+\cos^2\alpha$.

\begin{figure}
    \centering
    \includegraphics[width=0.8\linewidth]{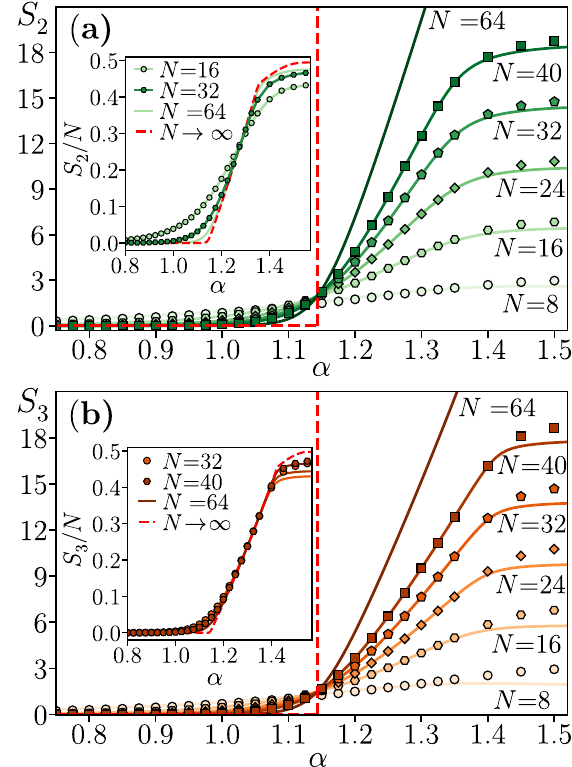}
    \caption{
Participation entropy of the decoded logical state for Clifford encoders with the postselected syndrome $\mathbf{s}=\mathbf{0}$ at a fixed code rate $r=k/N=1/2$. 
(a) The PE $\mathrm{S}_2$ as a function of the error strength $\alpha$. Solid lines represent the analytic formula~\eqref{eq:P2good}, while symbols denote the numerically obtained quenched averages of the PE. The inset shows the PE density $\mathrm{S}_2/N$ as a function of $\alpha$, with the red dashed line indicating the behavior in the thermodynamic limit ($N\to \infty$). (b) Analogous results for the PE $S_3$. Numerical data are computed using the Pauli propagation technique and averaged over more than 1000 circuit realizations.
    }
    \label{fig:entPart}
\end{figure}

As shown in previous studies of encoding-decoding dynamics~\cite{turkeshi2024error}, these quantities serve as secondary diagnostics for the transition. In the error-resilient phase ($\alpha<\alpha_c$), the successful recovery of a simple logical state implies vanishing entropy. In the error-vulnerable phase ($\alpha>\alpha_c$), the failure of decoding leaves the logical subsystem in a highly scrambled state, exhibiting volume-law entanglement and extensive participation entropy, indicative of multifractal delocalization in the Hilbert space.
We compare Eqs.~\eqref{eq:P2good}-~\eqref{eq:S3good} with the numerics for the quench PE in Fig.~\ref{fig:entPart}. Analogously for the fidelity computation, we observe quantitative agreement between our analytics and the numerics up to $N\le 40$ qubits.
This further highlights that, for the forced measurement regime, the physics is fully captured by the error-resilience phase transition.

\begin{figure}
    \centering
    \includegraphics[width=1\linewidth]{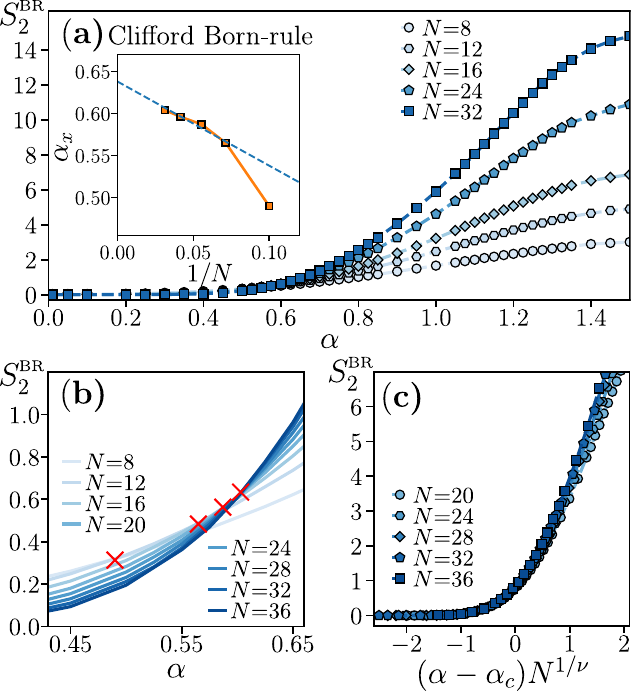}
    \caption{
Participation entropy  $\mathrm{S}^{\mathrm{BR}}_2$ of the decoded logical state for Clifford encoders with Born-rule syndrome averaging at a fixed code rate $r=k/N=1/2$. 
(a) The PE $\mathrm{S}^{\mathrm{BR}}_2$ as a function of the error strength $\alpha$. The inset shows the crossing point (displayed in panel (b)) as a function of $1/N$, alongside an extrapolation to the large-$N$ limit, where $\alpha_x \to 0.636 = \alpha_s \approx \alpha_c^{\mathrm{BR}}$. (c) The collapse of the PE when plotted as a function of the rescaled error strength $(\alpha-\alpha_s)N^{1/\nu}$, with the critical exponent $\nu=2$. Results are computed using the Pauli propagation technique and averaged over more than 1000 circuit realizations.
    }
    \label{fig:PEbr}
\end{figure}

The situation is drastically different when syndrome measurements are sampled by Born rule. Our large-scale exact numerical results are reported in Fig.~\ref{fig:PEbr} for PE $\mathrm{S}^{\mathrm{BR}}_2 \equiv \mathbb{E}_{U\sim \mathcal{E}}\sum_{\mathbf{s} } p_{U}(\mathbf{s}) S_2 (\rho_{L,U}(\mathbf{s}))$. As for the fidelity and stabilizer entropy, the participation entropy develops two competing phase transitions: the error-resilient one and the transition in the Born probability, which acts as a relevant perturbation and moves the critical point toward $\alpha_c\simeq 0.64$ with a critical exponent compatible with $\nu=2$. 

Overall, these findings confirms the narrative in the Main Text where two different type of transition occurs: post-selecting on syndrome measurement classes the physics is fully captured by the error-resilient transitions; in contrast, Born rule drive the system toward another universality class, which is not captured by the annealed average computations. 
In essence, this has analogy with measurement-induced phase transitions, where there is a spontaneous symmetry breaking in the replica statistical mechanics model, and the replica limit is key to capture the relevant physics~\cite{nahum2021measurementandentanglement,fava2023nonlinearsigmamodels,vasseur2019entanglementtransitionsfrom}.

\section{Boundary replica operators}
\label{app:bound}
This Appendix details the boundary replica operators for the quantities of interest in \textbf{Sec.}~\ref{sec:forced}.

\subsection{Normalization}
We start presenting the contribution for the annealed average denominators arising from the normalization condition in Eq.~\eqref{eq:pF_def}.
By the cyclicity of the trace, this can be immediately rewritten as 
\begin{equation}
p_{U}(\mathbf{s})
=
\Tr\left[\ket{\Psi_0} \!\bra{\Psi_0}\,U^\dagger V_\alpha^\dagger U\,\Pi_{I,\mathbf{s}}\,U^\dagger V_\alpha U\,\right]
\label{eq:mA1}
\end{equation}
A direct calculation shows that for any operators $A_1,A_2,A_3,A_4$,  the following  identity holds
\begin{equation}
\Tr\!\left[ A_1 A_2 A_3 A_4\right]
=
\Tr\!\left[ T_{(12)} (A_1 \otimes A_3)(A_2 \otimes A_4)\right],
\label{eq:A4}
\end{equation}
where the trace on the left-hand side is taken over $\mathcal{H}$, while the trace on the right-hand side is taken over $\mathcal{H}^{\otimes 2}$. Here $T_{(12)}$ is the representation of the transposition $\sigma=(12)$, i.e., the swap operator between the two replicas.
Since Eq.~\eqref{eq:mA1} is a product of four operators, we identify
$A_1=\ket{\Psi_0}\bra{\Psi_0}$,
$A_2= U^\dagger V_\alpha^\dagger U$,
$A_3=\Pi_{I,\mathbf{s}}$,
$A_4=U^\dagger V_\alpha U$.
Using \eqref{eq:A4} and regrouping terms, we obtain
\begin{equation}
p_{U}(\mathbf{s})
=
\Tr\!\left[ T_{(12)}  ( \ket{\Psi_0}\bra{\Psi_0} \otimes \Pi_{I,\mathbf{s}})  \,(U^{\dagger})^{\otimes 2} ( V_\alpha \otimes V_\alpha^\dagger) U^{\otimes 2}\right]. \nonumber
\label{eq:mA2}
\end{equation}
In the postselected case $\mathbf{s}=\mathbf{0}$, for the initial state $\ket{\Psi_0}=\ket{\mathbf{0}}_L\otimes \ket{\mathbf{0}}_S$ and
$\Pi_{I,\mathbf{0}} =I_L\otimes \ket{\mathbf{0}}\bra{\mathbf{0}}_S$,
this yields
\begin{align}
p_{U}(\mathbf{0})\nonumber
=\Tr\bigl[  T_{(12)} [\bigl(   |\mathbf{0}\rangle\langle\mathbf{0}|_L  
    \otimes  |\mathbf{0}\rangle\langle\mathbf{0}|_S\bigr)  \otimes (I_L \otimes & \ket{\mathbf{0}}\bra{\mathbf{0}}_S) ] \cdot &\\
\,(U^{\dagger})^{\otimes 2}  ( V_\alpha \otimes V_\alpha^\dagger)  U^{\otimes 2}\bigr]
\label{eq:mA3}
\end{align}
From this expression we can directly read off the boundary operator $\mathcal{B}_{\mathrm{den}}^{F}\equiv \mathcal{B}(I)$ as well as the form of $A_U^{(2)}$ in Eq.~\eqref{eq:A2_def}.
Fluctuations of the fidelity (or of the stabilizer R\'enyi entropy) involve multiple copies of the logical state and require evaluating averages of $p_U^{n}(\mathbf{0})$ for integer $n\ge 2$.
Let us consider $n=2$ for concreteness. We have
\begin{equation}
\begin{split}
	p_U^2(\mathbf{0}) &=\Tr[T_{(12)}\mathcal{B}(I) A_U^{(2)}]^2 \\
	&= \Tr\left[(T_{(12)}\mathcal{B}(I) A_U^{(2)})^{\otimes 2}\right] \\&= \Tr\left[T_{(12)(34)} (\mathcal{B}(I))^{\otimes 2} A_U^{(4)}\right]\label{eq:zionnnn1}
\end{split}
\end{equation}
where in the second step we used $\Tr[A]\Tr[B]=\Tr[A\otimes B]$, while in the last step the natural representation embedding of $\mathrm{S}_2\times \mathrm{S}_2\mapsto \mathrm{S}_4$ with $T_{(12)}^{\otimes 2} \mapsto T_{(12)(34)}\equiv T_{\sigma_4}$ and $(A_U^{(2)})^{\otimes 2}\equiv A_U^{(4)}$. 
With a similar argument, for any integer $n\geq 1$ we have
\begin{equation}
	p_U^n(\mathbf{0}) = \Tr\left[T_{\sigma_{2n}} (\mathcal{B}(I))^{\otimes n} A_U^{(2n)}\right],
\end{equation}
where $\sigma_{2n}=(12)\cdots(2n-1\ 2n)$.
In the next subsections we present the boundary operators required for the numerators in the replica computations.

\subsection{Fidelity and fluctuations}
\label{app:fideB}
The replica representation for the fidelity numerator $m_{F,U}(\mathbf{s})$ (cf. Eq.~\eqref{eq:mF_def}) follows exactly the same steps as for the normalization factor.
The only modification is the replacement $\Pi_{I,\mathbf{s}} \mapsto \Pi_{\psi,\mathbf{s}}$.
For the initial logical state $\ket{\psi}_L=\ket{\mathbf{0}}_L$ and postselected syndrome $\mathbf{s}=\mathbf{0}$, this yields the boundary operator quoted in Eq.~\eqref{eq:fidB}.
Likewise, the second moment $n=2$ of the numerator, $\mathbb{E}_{U\sim \mathcal{E}}\!\left[m_{F,U}^2\right]$, is obtained by the same embedding used in Eq.~\eqref{eq:zionnnn1} and gives
$B^{m_F}=\mathcal{B}(|\mathbf{0}\rangle\langle\mathbf{0}|)^{\otimes 2}$.

\subsection{Stabilizer R\'{e}nyi entropy }

To construct the replica boundary operator entering the numerator of the annealed SRE $m^{(q)}_{M,U}(\mathbf{s})$ [cf. Eq.~\eqref{eq:sreAN3}], we note it is a linear combination of terms of the same form as $p_U^{2q}(\mathbf{s})$, with the replacement $\Pi_{I,\mathbf{s}}\mapsto \Pi_{P,\mathbf{s}}$. For $\mathbf{s}=\mathbf{0}$ this gives
\begin{align}
m^{(q)}_{M,U}&(\mathbf{0})
=\frac{1}{2^k} \sum_{P \in \mathcal{P}_k }\Tr\left[  T_{\sigma_{4q}} 
[   |\mathbf{0}\rangle\langle\mathbf{0}|   \otimes \Pi_{P,\mathbf{0}} ]^{\otimes{2q}} 
\,A_U^{(4q)}\right]\nonumber \\
&=\frac{1}{2^k}\sum_{P\in \mathcal{P}_k}\Tr\left[  T_{\sigma_{4q}}\mathcal{B}(P)^{\otimes 2q} A_U^{(4q)}\right]
\label{eq:mA9}
\end{align}
which, by the linearity of the trace, fixes the form of the replica boundary operator $\mathcal{B}^{(M_q)}_{\mathrm{num}}$ in Eq.~\eqref{eq:B8_def}.

\subsection{Participation entropy}
The replica boundary operator in the numerator for the annealed participation entropy in Eq.~\eqref{eq:PEann} parallels that of the SRE. In this case, $m^{(q)}_{S,U}(\mathbf{s})$ is a linear combination of terms of the form $p_U^{2q}(\mathbf{s})$, with the replacement $\Pi_{I,\mathbf{s}}\mapsto \Pi_{|\mathbf{x}\rangle\langle \mathbf{x}|,\mathbf{s}}$. For $\mathbf{s}=\mathbf{0}$ this gives
\begin{align}
m^{(q)}_{S,U}(\mathbf{0})
&= \sum_{\mathbf{x} \in \mathbb{Z}_2^{k} }\operatorname{Tr}\Bigl[ T_{\sigma_{4q}}
\bigl[ |\mathbf{0}\rangle\langle\mathbf{0}| \otimes \Pi_{|\mathbf{x}\rangle\langle \mathbf{x}|,\mathbf{0}} \bigr]^{\otimes{2q}}
\,A_U^{(4q)}\Bigr]\nonumber \\
&=\sum_{\mathbf{x} \in \mathbb{Z}_2^{k} }\operatorname{Tr}\Bigl[ T_{\sigma_{4q}}\mathcal{B}(|\mathbf{x}\rangle\langle \mathbf{x}|)^{\otimes 2q} A_U^{(4q)}\Bigr],
\label{eq:mA92}
\end{align}
where we read by linearity the expression of $\mathcal{B}^{(M_q)}_{\mathrm{num}}$ in Eq.~\eqref{eq:srePE2}.

\subsection{Simplification to replica qubit problems}
\label{app:onsite}
The annealed averages can be evaluated efficiently because the replica boundary operators factorize across physical qubits. Concretely, they can be written as tensor products of $N$ single-qubit operators acting on certain $q$ replicas. 
As a representative example, consider the numerator of the second annealed SRE (requiring $8$ replicas)
\begin{align}
\mathcal{B}^{(M_2)}_{\mathrm{num}} =& T_{\sigma_8} \!
\sum_{P\in{\mathcal{P}}_k}\frac{\bigl[\mathcal{B}(P)\bigr]^{\otimes 4}}{2^k} =
\nonumber\\
=&\frac{ t_{\sigma_8}^{\otimes N}}{2^k}\sum_{P\in{\mathcal{P}}_k}[|\mathbf{0}\rangle\langle\mathbf{0}|_L  \otimes |\mathbf{0}\rangle\langle\mathbf{0}|_S  \otimes P_L \otimes |\mathbf{0}\rangle\langle\mathbf{0}|_S ]^{\otimes{4}} =
\nonumber\\
=& [ t_{\sigma_8}  Q^M_4]^{\otimes k} \otimes [ t_{\sigma_8} (|0\rangle\langle0|)^{\otimes 8} ]^{\otimes (N-k)},
\label{eq:Bprod}
\end{align}
Here we used the factorization of the permutation representation on $N$ qubits,
$T_{\sigma}=\bigotimes_{i=1}^N t_{\sigma}^{(i)}$, and then grouped the resulting $8$-copy operators according to whether they act on one of the $k$ logical sites or on one of the $N-k$ ancillary sites. 
The nontrivial single-site object on the logical sites is the replica-qubit operator
\begin{equation}
Q^M_n \equiv \frac{1}{2}\sum_{P\in\mathcal{P}_1}\bigl(|0\rangle\langle 0|\otimes P\bigr)^{\otimes 2n},
\end{equation}
which for $n=2$ reproduces operator in Eq.~\eqref{eq:Bprod}.
A similar simplification holds for other observables; only the definition of the replica-qubit operator $Q_n$ changes. For instance, for the numerator of the second annealed average participation entropy involves $Q^{S}_2$, where
$Q^{S}_n \equiv \sum_{x\in \{0,1\}} (|0\rangle\langle 0|\otimes |x\rangle\langle x|)^{\otimes n}$.
In all other cases considered throughout this manuscript, the reduction to replica-qubit problems follows directly from the tensor-product structure of the replica boundary operators. 
Finally, the permutation operators $T_{\sigma}$ (and, in the Clifford case, the corresponding elements of the commutant entering the Schur--Weyl expansion) factorize in the same single-site manner. As a result, all annealed quantities reduce to products of single-qubit replica traces, which can be efficiently evaluated symbolically.

\section{Averages over the Unitary and Clifford groups}
\label{App:averages}

In this Appendix, we detail the framework for computing group averages of operators acting on the $n$-copy Hilbert space $\mathcal{H}^{\otimes n}$.
This framework is required for the calculations of the ensemble averages of the operator $A_U^{(n)}=(U^\dagger)^{\otimes n} (V_\alpha\otimes V_\alpha^\dagger)^{\otimes (n/2)} U^{\otimes n}$, entering all the main computations of the manuscript where $n$ is an even number. 
Relevant cases are $n=2q$ for $q=1,2,3$, encompassing the computation of the fidelity and its fluctuation [cf. \textbf{Sec.}~\ref{subsec:fidelity}], as well as the participation entropy [cf. \textbf{App.}~\ref{app:parti}], and $n=4q$ with $q=2,3$ for the computation of the stabilizer entropy [cf. \textbf{Sec.}~\ref{subsec:magica}].
We refer to Ref.~\cite{Mele2024introductiontohaar} and Ref.~\cite{Bittel25commutant} for a pedagogical introduction to the averages over the unitary and Clifford groups, respectively.

\subsection{Unitary Group Averages}
Consider an operator $O$ acting on $\mathcal{H}^{\otimes n}$. We define the average over the full unitary group $\mathcal{U}(2^N)$ as the integral
\begin{equation}
\mathbb{E}_{U \sim \mathcal{U}(2^N)} \left[ (U^\dagger)^{\otimes n} O U^{\otimes n} \right]
= \int_{\mathcal{U}(2^N)} d\mu(U) \, (U^\dagger)^{\otimes n} O U^{\otimes n}.
\label{eq:HaarTwirl}
\end{equation}
By Schur-Weyl duality, the commutant~\footnote{The commutant of a set of operators $\mathcal{S}$ is the algebra of all operators that commute with every element in $\mathcal{S}$. In the context of unitary (or Clifford) groups, it refers to the set of operators $X$ on $\mathcal{H}^{\otimes n}$ satisfying $[X, U^{\otimes n}] = 0$ for all unitary (or Clifford) operators.} of the unitary group of $\mathcal{H}^{\otimes n}$ is spanned by the permutation operators $T_\pi$ (with $\pi \in \mathrm{S}_{n}$), which permute the $n$ tensor factors. 
Consequently, the Haar average projects $O$ onto the span of these permutations:
\begin{equation}
    \mathbb{E}_{U \sim \mathcal{U}(2^N)} \left[ (U^\dagger)^{\otimes n} O U^{\otimes n} \right]
    = \sum_{\pi, \sigma \in \mathrm{S}_{n}} \mathrm{Wg}_{\pi,\sigma} \Tr(T_\sigma O) T_\pi,
    \label{eq:UnitaryWeingarten}
\end{equation}
where $\mathrm{Wg}_{\pi,\sigma}$ denote the elements of the Weingarten matrix. This matrix is obtained as the inverse (or pseudo-inverse for $2^N < n$) of the Gram matrix of permutation operators, defined by $G_{\pi, \sigma} = \Tr(T_\pi^\dagger T_\sigma)$. The Weingarten functions admit a closed-form representation in terms of a sum over integer partitions $\lambda \vdash n$~\cite{Collins2006}, which we utilize for calculations involving $n=2,4$ replicas.
For a large Hilbert space dimension $D=2^N$, the Weingarten function admits an asymptotic expansion in powers of $1/D$. The leading order term is given by $\text{Wg}_{\pi,\sigma}\approx D^{-n}\delta_{\pi,\sigma} + O(D^{-n-1})$~\cite{Collins2006}. Retaining only the leading contribution ($\sigma = \pi$) in Eq.~\eqref{eq:UnitaryWeingarten} yields the \textit{diagonal approximation}~\cite{Schuster2025,dowling2025freeindependenceunitarydesign}. This approximation introduces errors that decay exponentially with the system size $N$, and we employ it to evaluate the SRE averages involving $n=8$ and $n=12$ replicas.

\subsection{Clifford Group Averages}
\label{app:cliffINT}
The average of an operator $O$ acting on $\mathcal{H}^{\otimes n}$ over the Clifford group $\mathcal{C}_N$ is a discrete sum 
\begin{equation}
\mathbb{E}_{C \sim \mathcal{C}_N} \left[ (C^\dagger)^{\otimes n} O C^{\otimes n} \right]
    = \frac{1}{|\mathcal{C}_n|} \sum_{C \in \mathcal{C}_n} (C^\dagger)^{\otimes n} O C^{\otimes n}.
    \label{eq:CliffordTwirl}
\end{equation}
For qubits, the Clifford group forms a unitary 3-design~\cite{Gross07design,Dankert09,zhu2017multiqubitcliffordgroups,webb2016clifford}, implying that for $n \leq 3$, the Clifford average is identical to the unitary average: $\mathbb{E}_{C \sim \mathcal{C}_N} \left[ (C^\dagger)^{\otimes n} O C^{\otimes n} \right]=\mathbb{E}_{U \sim \mathcal{U}(2^N)} \left[ (U^\dagger)^{\otimes n} O U^{\otimes n} \right]$.

For $n \geq 4$, the Clifford commutant is strictly larger than the span of permutation operators, and the associated operators can be parametrized using stochastic Lagrangian subspaces~\cite{Gross2021}. 
In the following discussion, and in our symbolic implementation~\cite{OURzenodo}, we employ an alternative parameterization of the Clifford commutant based on the set of reduced Pauli monomials
$\mathfrak{P}_n = \{ \Omega \}$~\cite{Bittel25commutant}. This formulation expresses the Clifford average in terms of the operators $\Omega\in \mathfrak{P}_n$ as
\begin{equation}
\mathbb{E}_{C \sim \mathcal{C}_N} \left[ (C^\dagger)^{\otimes n} O C^{\otimes n} \right]
    =\sum_{\Omega,\Omega' \in \mathfrak{P}_n} \mathrm{Wg}^{\mathcal{C}_N}_{\Omega,\Omega'} \Tr(\Omega O)  {\Omega'},
    \label{eq:CliffordTwirl2}
\end{equation}
where $\text{Wg}^{\mathcal{C}_N}_{\Omega,\Omega'}$ are Clifford-Weingarten functions, which are be obtained as the pseudo-inverse of the Gram matrix $G^{\mathcal{C}_N}_{\Omega, {\Omega'}} = \Tr(\Omega^\dagger {\Omega'})$.
For a system of $N$ qubits, the cardinality of $\mathfrak{P}_n$, which corresponds to the dimension of the $n$-th Clifford commutant, is given, for $N \geq n-1$, by~\cite{Magni2025quantumcomplexity}:
\begin{equation}
    |\mathfrak{P}_n| = \prod_{m=0}^{n-2} (2^m + 1),
    \label{eq:clicomdim}
\end{equation}
and the marginal sum of the Clifford-Weingarten matrix is constant:
\begin{equation}
    \sum_{\Omega \in \mathfrak{P}_n}  \mathrm{Wg}^{\mathcal{C}_N}_{\Omega,\Omega'} = 2^{-2nN} \prod_{m=0}^{n-2} \left(1 + 2^{m-2N}\right)^{-1}.
\end{equation}
These identities serve as useful consistency checks during the generation of $\mathfrak{P}_n$. 
Exact computation can be performed up to $n=4$ copy computations. For the participation entropy $S_3$ and the stabilizer entropy $M_2$ we require reaching larger replica order $n$. 
In these scenarios, we use the asymptotics of the Clifford-Weingarten function for large $N$, $\mathrm{Wg}^{\mathcal{C}_N}_{\Omega,\Omega'} \approx D^{-n}\delta_{\Omega,\Omega'} + O(D^{-n-1})$ where $D=2^N$. This allows for a diagonal approximation similar to the unitary case, which we utilize for calculations involving $n=6$ and $n=8$ replicas.

The Clifford average in Eq.~\eqref{eq:CliffordTwirl2} shares a similar structure with the unitary average in Eq.~\eqref{eq:UnitaryWeingarten}. Since the elements of $\mathfrak{P}_n$ factorize as tensor products of $n$-replica single-qubit operators, the calculation of annealed averages in the Clifford case follows a pattern analogous to the unitary case, provided the basis set $\mathfrak{P}_n$ is correctly generated. The set $\mathfrak{P}_n$ is constructed as the unique operators of the form 
\begin{equation}
    \Omega = T_{\pi}\Omega_V T_{\sigma},
    \label{eq:omegaOP}
\end{equation}
where $\Omega_V$ are \emph{primitive Pauli monomials} (whose number and form depend on $k$) and $T_\pi$ are the permutation operators.

The primitive Pauli monomials are specified by a set $V = \{ \mathbf{b}^1, \dots, \mathbf{b}^m \}$ of $m$ bitstrings $\mathbf{b}^\ell\in \mathbb{Z}_2^k$ such that $\sum_{j=1}^n b_j^\ell=0\mod 2$, as
\begin{equation}
\Omega_V = \prod_{\ell=1}^m \left( \frac{1}{2^N} \sum_{P \in \mathcal{P}_N} \bigotimes_{j=1}^k P^{b^{\ell}_j} \right).
\label{eq:Reduced}
\end{equation}
In the following, we detail how to construct these operators in $n=4,6,8$ replicas. 

\subsubsection{$n=4$ replicas}

For the case of $n=4$ replicas, the dimension of the Clifford commutant for a system of $N$ qubits ($N \geq 3$) is $|\mathfrak{P}_4|= 30$, strictly larger than the unitary commutant spanned by the permutation operators, which has dimension $|\mathrm{S}_4| = 24$.

Besides the $24$ permutation operators $T_\sigma$ obtained from~\eqref{eq:omegaOP} setting $\Omega_\emptyset$, the additional Pauli monomials are generated by 
\begin{equation}
    \Omega_{\{{(1,1,1,1)}\}} = \frac{1}{2^N} \sum_{P \in \mathcal{P}_N} P^{\otimes 4}.
\end{equation}
Left and right action of permutation operators on $\Omega_{\{{(1,1,1,1)}\}}$ yields $6$ unique operators $\Omega$ which complete the basis of $4$-replica Clifford commutant, as summarized Table~\ref{tab:k4basis}.
\begin{table}[h!]
\centering
\renewcommand{\arraystretch}{1.3} 
\begin{tabular}{|l|c|c|}
\hline
\textbf{Class Name} & \textbf{Generator Basis } $V$ & \textbf{Sector Size} \\ \hline
Permutations & $\emptyset$ & 24 \\ \hline
$\Omega^{(4)}$ & $\{ (1,1,1,1) \}$ & 6 \\ \hline
\multicolumn{2}{|r|}{{Total Dimension} $|\mathfrak{P}_4|$} & {30} \\ \hline
\end{tabular}
\caption{Structure of the $k=4$ Clifford commutant.}
\label{tab:k4basis}
\end{table}

To evaluate the Clifford average efficiently, we perform a symbolic computation of the Clifford-Weingarten functions $\mathrm{Wg}^{\mathcal{C}_N}$. This requires inverting the $30 \times 30$ Gram matrix $G^{\mathcal{C}_N}$. We leverage the known structure of the unitary Weingarten calculus to avoid computing the full matrix inverse from scratch. The Gram matrix naturally admits a block structure induced by the decomposition of the commutant into the span of permutation operators and the non-permutation sector spanned by the Pauli monomials generated by $\Omega_{\{{(1,1,1,1)}\}}$. To invert this matrix symbolically, we employ the Schur complement method, which allows us to express the full inverse in terms of the known unitary Weingarten matrix (the inverse of the permutation block) and the inverse of the Schur complement. The resulting Clifford-Weingarten functions $\mathrm{Wg}^{\mathcal{C}_N}_{\Omega,\Omega'}$ are available at~\cite{OURzenodo}.

\subsubsection{$k=6$ replicas}
For $k=6$, the Clifford commutant decomposes into the standard permutation sector and three additional sectors generated by reduced Pauli monomials. These sectors are specified by the operators $\Omega_V$, which are constructed from the four sets of generator bitstrings $V \subseteq \mathbb{Z}_2^6$ listed in Table~\ref{tab:k6basis} using Eq.~\eqref{eq:Reduced} with $n=6$. 
The full set  $\mathfrak{P}_6$, i.e., the basis of the $n=6$ Clifford commutant is obtained by closing this set under the left and right action of the permutation group $\mathrm{S}_6$, cf. Eq.~\eqref{eq:omegaOP}. This procedure yields a total of $4590$ unique basis elements, in agreement with the dimension formula in Eq.~\eqref{eq:clicomdim}.

\begin{table}[h!]
\centering
\renewcommand{\arraystretch}{1.3} 
\begin{tabular}{|l|c|c|}
\hline
\textbf{Class Name} & \textbf{Generator Basis } $V$ & \textbf{Sector Size} \\ \hline
Permutations & $\emptyset$ & 720 \\ \hline
$\Omega^{(6)}$ & $\{ (1,1,1,1,1,1) \}$ & 720 \\ \hline
$\Omega^{(4)}$ & $\{ (1,1,1,1,0,0) \}$ & 2700 \\ \hline
$\Omega^{(4,4)}$ & 
\begin{tabular}{@{}c@{}}
    $\{ (1,1,1,1,0,0), $ \\ 
    $ \phantom{\{} (0,0,1,1,1,1) \}$
\end{tabular} 
& 450 \\ \hline
\multicolumn{2}{|r|}{{Total Dimension} $|\mathfrak{P}_6|$} & {4590} \\ \hline
\end{tabular}
\caption{Structure of the $k=6$ Clifford commutant. }
\label{tab:k6basis}
\end{table}

\subsubsection{$k=8$ replicas}

For the case of $k=8$ replicas, the Clifford commutant exhibits a vast increase in complexity, yielding a total dimension of $|\mathfrak{P}_8| = 9845550$. This basis partitions into $13$ distinct classes of operators $\Omega_V$~\cite{Bittel25commutant}, which are generated by specific sets of bitstrings $V \subseteq \mathbb{Z}_2^8$ via Eq.~\eqref{eq:Reduced}. The full set of unique basis elements is recovered by applying the left and right action of the symmetric group $\mathrm{S}_8$ to these generators, the structure of which is summarized in Table~\ref{tab:k8basis}.

\begin{table}[h!]
\centering
\resizebox{\columnwidth}{!}{
\begin{tabular}{|l|c|c|}
\hline
\textbf{Class Name} & \textbf{Generator Basis } $V$ & \textbf{Sector Size} \\ \hline
Permutations & $\emptyset$ & 40320 \\ \hline
$\Omega^{(8)}$ & $\{ (1,1,1,1,1,1,1,1) \}$ & 40320 \\ \hline
$\Omega^{(6)}$ & $\{ (1,1,1,1,1,1,0,0) \}$ & 1128960 \\ \hline
$\Omega^{(4)}$ & $\{ (1,1,1,1,0,0,0,0) \}$ & 705600 \\ \hline
$\Omega^{(4,4)}_{\text{disj}}$ & 
\begin{tabular}{@{}c@{}}
    $\{ (1,1,1,1,0,0,0,0), $ \\ 
    $ \phantom{\{} (0,0,0,0,1,1,1,1) \}$
\end{tabular} 
& 88200 \\ \hline
$\Omega^{(4,4)}_{\text{ov1}}$ & 
\begin{tabular}{@{}c@{}}
    $\{ (1,1,1,1,0,0,0,0), $ \\ 
    $ \phantom{\{} (0,0,0,1,1,1,1,0) \}$
\end{tabular} 
& 2822400 \\ \hline
$\Omega^{(4,4)}_{\text{ov2}}$ & 
\begin{tabular}{@{}c@{}}
    $\{ (1,1,1,1,0,0,0,0), $ \\ 
    $ \phantom{\{} (0,0,1,1,1,1,0,0) \}$
\end{tabular} 
& 705600 \\ \hline
$\Omega^{(4,6)}$ & 
\begin{tabular}{@{}c@{}}
    $\{ (1,1,1,1,0,0,0,0), $ \\ 
    $ \phantom{\{} (0,0,1,1,1,1,1,1) \}$
\end{tabular} 
& 2116800 \\ \hline
$\Omega^{(6,6)}$ & 
\begin{tabular}{@{}c@{}}
    $\{ (1,1,1,1,1,1,0,0), $ \\ 
    $ \phantom{\{} (0,0,1,1,1,1,1,1) \}$
\end{tabular} 
& 1411200 \\ \hline
$\Omega^{(3)}_{\text{chain}}$ & 
\begin{tabular}{@{}c@{}}
    $\{ (1,1,1,1,0,0,0,0), $ \\ 
    $ \phantom{\{} (0,0,0,0,1,1,1,1), $ \\
    $ \phantom{\{} (0,0,1,1,1,1,0,0) \}$
\end{tabular} 
& 22050 \\ \hline
$\Omega^{(3)}_{\text{star}}$ & 
\begin{tabular}{@{}c@{}}
    $\{ (1,1,1,1,0,0,0,0), $ \\ 
    $ \phantom{\{} (0,0,1,1,1,1,0,0), $ \\
    $ \phantom{\{} (0,1,0,1,1,0,1,0) \}$
\end{tabular} 
& 57600 \\ \hline
$\Omega^{(3)}_{\text{cyc}}$ & 
\begin{tabular}{@{}c@{}}
    $\{ (1,1,1,1,0,0,0,0), $ \\ 
    $ \phantom{\{} (0,0,1,1,1,1,0,0), $ \\
    $ \phantom{\{} (1,0,1,0,0,0,1,1) \}$
\end{tabular} 
& 705600 \\ \hline
$\Omega_{\text{Hamming}}$ & 
\begin{tabular}{@{}c@{}}
    $\{ (1,1,1,1,0,0,0,0), $ \\ 
    $ \phantom{\{} (0,0,1,1,1,1,0,0), $ \\
    $ \phantom{\{} (0,0,1,1,0,0,1,1), $ \\
    $ \phantom{\{} (1,0,1,0,1,0,1,0) \}$
\end{tabular} 
& 900 \\ \hline
\multicolumn{2}{|r|}{{Total Dimension} $|\mathfrak{P}_8|$} & {9845550} \\ \hline
\end{tabular}
}
\caption{Structure of the $n=8$ Clifford commutant.}
\label{tab:k8basis}
\end{table}

\subsubsection{Generation of the Clifford commutant basis}

To set up the Clifford commutant bases at $n=4,6,8$ replicas, we generate the sets of unique operators of the form
\begin{equation}
\mathfrak{P}_k =\text{unique}\left\{ { T_{\pi}\Omega_V T_{\sigma}  \mid \pi, \sigma \in \mathrm{S}_n } \right \},
\label{eq:omegaOP2}
\end{equation}
where $\Omega_V$ runs over the classes of Pauli monomials at given $n$.
The naive cardinality of the search space is $|\mathrm{S}_n|^2$ (e.g., for $n=8$, $(40320)^2 \approx 1.6 \times 10^9$), which renders brute-force generation computationally infeasible.

To efficiently generate the basis $\mathfrak{P}_n$, we exploit the group structure of $\mathrm{S}_n$ and the sparsity of the seed operators $\Omega_V$. We utilize the identity that the set generated by $(T_\pi \Omega_V T_\pi^\dagger) T_\sigma$ is identical to that of $T_\pi \Omega_V T_\sigma$, as the inner $T_\pi^\dagger$ is absorbed by the arbitrary choice of the right-multiplying permutation $T_\sigma$.

The algorithm proceeds in two phases:
\begin{enumerate}
  \item For each generator class $\Omega_V$, we compute the set of conjugation representatives:
\begin{equation}
    \mathcal{L}_V = \text{unique}\left( \{ T_\pi \Omega_V T_\pi^\dagger \mid \pi \in \mathrm{S}_n \} \right).
\end{equation}
Due to the internal symmetries of $\Omega_V$, the size of $\mathcal{L}_V$ is significantly smaller than $|\mathrm{S}_n|$. This "collapse" is particularly pronounced at $n=8$ replicas, where the largest set $\mathcal{L}_V$ (corresponding to the class $\Omega^{(3)}_{\text{cyc}}$) consists of only $2520$ elements, a fraction of the permutation group size.

\item  For each unique representative $L \in \mathcal{L}_V$, we generate the full orbit under right-multiplication by all elements of $\mathrm{S}_n$
\begin{equation}
    \mathcal{M}_{L} = \{ L T_\sigma \mid \sigma \in \mathrm{S}_n \}.
\end{equation}
\end{enumerate}
It is crucial to note that while the representatives $L \in \mathcal{L}_V$ are distinct under conjugation, their right-multiplication orbits $\mathcal{M}_L$ are not guaranteed to be disjoint. A collision occurs when two distinct representatives $L, L' \in \mathcal{L}_V$ map to the same operator, i.e., 
     $L T_\sigma = L' T_\pi$
for some permutations $\sigma, \pi \in \mathrm{S}_n$. 
Such collisions occur at $n=8$ replicas.
Consequently, the final basis is formed by the union of these partial orbits across all classes, filtered for uniqueness:
\begin{equation}
 \mathfrak{P}_n = \text{unique}\left( \bigcup_{V} \bigcup_{L \in \mathcal{L}_V} \mathcal{M}_{L} \right).
\end{equation}
These methods, implemented in a symbolic algebra package and available at~\cite{OURzenodo}, enable calculation of the annealed averages for the Clifford encoders.

\section{Heuristic bound on magic resources in the error-protecting phase}
\label{app:heuristic_magic_bound}

A key message of this work is that, while Haar and Clifford encoders differ markedly at the level of higher moments, the onset of magic in encoding–decoding circuits in the post-selected $\mathbf{s}=\mathbf{0}$ case, is ultimately controlled by the same physical mechanism: the error-resilience transition. 
In this limit, the transition in stabilizer Rényi entropy coincides with the logical-fidelity transition: Below the critical error strength $\alpha_c$, the decoding succeeds with unit fidelity, and any magic injected by the coherent error layer is necessarily expelled from the logical subspace.

In this Appendix, we provide a simple heuristic bound that makes this statement explicit for Haar and Clifford encoders. The argument complements the replica-based analysis of \textbf{Sec.}~\ref{eq:forzaitalia} and \textbf{Sec.}~\ref{subsec:maclif}.

Fix the syndrome outcome $\mathbf{s}=\mathbf{0}$ and denote the corresponding decoded logical pure state by
$\ket{\psi_U(\mathbf{0})}_L$ [cf. Eq.~\eqref{eq:stateS1}]. For notational simplicity, we write $\ket{\psi}\equiv\ket{\psi_U(\mathbf{0})}_L$ and take the initial (and target) logical state to be $\ket{\mathbf{0}}_L\equiv\ket{0}^{\otimes k}$. 
The (conditional) logical fidelity is $F \equiv \bigl|\langle \mathbf{0}|\psi\rangle\bigr|^2$. 
In the error-protecting phase ($\alpha<\alpha_c$), the forced-measurement protocol ensures $F\to 1$ in the thermodynamic limit. For $F>1/2$, we may decompose the state as
\begin{equation}
|\psi\rangle
=
\sqrt{F}\,|\mathbf{0}\rangle
+
\sqrt{1-F}\,e^{i\theta}\,|\phi\rangle,
\qquad
\langle \mathbf{0}|\phi\rangle = 0 .
\end{equation}

The stabilizer Rényi entropy $M_2$ involves a sum over all Pauli strings $P\in\mathcal{P}_k$. Since all terms in the sum are non-negative, restricting to any subset $\mathcal{S}\subset\mathcal{P}_k$ yields a lower bound on the argument of the logarithm:
\begin{equation}
\begin{split}
\frac{1}{2^k}\sum_{P\in\mathcal{P}_k}\langle P\rangle_\psi^4
\;&\ge\;
\frac{1}{2^k}\sum_{P\in\mathcal{S}}\langle P\rangle_\psi^4,
\\
\langle P\rangle_\psi &\equiv \langle\psi|P|\psi\rangle .
\end{split}
\end{equation}
We choose $\mathcal{S}$ to be the stabilizer group of $\ket{\mathbf{0}}$,
\begin{equation}
\mathcal{S}=\mathrm{Stab}(\ket{\mathbf{0}})
=
\{Z_1^{s_1}\cdots Z_k^{s_k}\;|\; s_i\in\{0,1\}\},
\end{equation}
which has cardinality $|\mathcal{S}|=2^k$.

For any $S\in\mathcal{S}$, define the projector onto its $-1$ eigenspace,
\begin{equation}
\Pi_-=\frac{I-S}{2}.
\end{equation}
Since $S|\mathbf{0}\rangle=|\mathbf{0}\rangle$, we have $\Pi_-|\mathbf{0}\rangle=0$. Using the above decomposition of $\ket{\psi}$, the expectation value of $S$ is
\begin{align}
\langle S\rangle_\psi
&=
\langle\psi|(I-2\Pi_-)|\psi\rangle
=1-2\langle\psi|\Pi_-|\psi\rangle \nonumber\\
&=
1-2(1-F)\langle\phi|\Pi_-|\phi\rangle
\;\ge\; 1-2(1-F)=2F-1 ,
\end{align}
where we used $0\le\langle\phi|\Pi_-|\phi\rangle\le 1$.

For $F>1/2$, this bound is strictly positive, and hence
$\langle S\rangle_\psi^4 \ge (2F-1)^4$ for all $S\in\mathcal{S}$. Summing over the stabilizer group gives
\begin{equation}
\frac{1}{2^k}\sum_{S\in\mathcal{S}}\langle S\rangle_\psi^4
\;\ge\;
(2F-1)^4 .
\end{equation}

Inserting this bound into the definition of the stabilizer Rényi entropy yields
\begin{align}
M_2(|\psi\rangle)
&=
-\log_2\!\left(
\frac{1}{2^k}\sum_{P\in\mathcal{P}_k}\langle P\rangle_\psi^4
\right)
\;\le\;
-\log_2\!\big[(2F-1)^4\big] \nonumber\\
&=
-4\log_2(2F-1),
\qquad (F>1/2).
\end{align}
To make contact with the error-resilient fixed point, we expand for high fidelity. Writing $F=1-\delta$ with $\delta\ll 1$, we find
\begin{equation}
M_2(|\psi\rangle)
\;\le\;
-4\log_2(1-2\delta)
\simeq
\frac{8}{\ln 2}\,\delta
+
O(\delta^2).
\end{equation}

In the forced-measurement ensemble, the error-protecting phase is characterized by $F \to 1$ for the trivial syndrome $\mathbf{s}=\mathbf{0}$. The bound above then implies that the stabilizer Rényi entropy must vanish continuously, $M_2 \to 0$, in this regime. In other words, as long as decoding succeeds with unit fidelity, the logical state is forced arbitrarily close to a stabilizer state, and any magic injected by the coherent error layer is completely expelled from the logical subspace. 
The same conclusion holds for Clifford encoders in the nontrivial syndrome sectors $\ell_{\mathbf{s}\neq\mathbf{0}}$, where the fidelity likewise approaches unity throughout the error-resilient phase. This establishes a direct and physically transparent link between error resilience and the absence of magic in the forced-measurement phase, fully consistent with the exact replica analysis presented in the main text.


%

\end{document}